\documentclass[
	superscriptaddress,
	showkeys,
	aps,
	pra,
    10pt,
	longbibliography,
	floatfix,
	nofootinbib
]{revtex4-2}

\usepackage[hyperref]{xcolor}
	\definecolor{goethe-blau}{cmyk}{1.0,0.2,0.0,0.4}
	\definecolor{hellgrau}{cmyk}{0.04,0.04,0.05,0.02}
	\definecolor{sandgrau}{cmyk}{0.12,0.09,0.13,0.0}
	\definecolor{dunkelgrau}{cmyk}{0.25,0.25,0.30,0.75}
	\definecolor{emo-rot}{cmyk}{0.04,1.0,0.8,0.07}
	\definecolor{purple}{cmyk}{0.08,1.0,0.3,0.36}
	\definecolor{senfgelb}{cmyk}{0.01,0.25,1.0,0.05}
	\definecolor{gruen}{cmyk}{0.62,0.4,0.87,0.09}
	\definecolor{magenta}{cmyk}{0.08,0.86,0.12,0.12}
	\definecolor{orange}{cmyk}{0.0,0.7,1.0,0.04}
	\definecolor{sonnengelb}{cmyk}{0.0,0.12,0.95,0.0}
	\definecolor{helles-gruen}{cmyk}{0.4,0.17,0.81,0.07}
	\definecolor{lichtblau}{cmyk}{0.8,0.0,0.06,0.04}

\usepackage[
colorlinks=true,
linkcolor=goethe-blau,
citecolor=goethe-blau,
urlcolor=goethe-blau,
breaklinks,
pdfstartview=FitH,
bookmarksopenlevel=0,
bookmarks=true,
pdftoolbar=true,
pdfauthor={{Fabrizio Murgana, Adrian Koenigstein, Dirk H. Rischke}}
pdftitle={Reanalysis of critical exponents for the O(N) model via a hydrodynamic approach to the Functional Renormalization Group},
pdfkeywords={FRG, fluid dynamics, critical exponents}
]{hyperref}

\newcommand{\av}[1]{\langle #1 \rangle}
\newcommand{\ave}[1]{\langle #1 \rangle}

\usepackage{verbatim}
\usepackage{amsmath}
\usepackage{graphicx}
\usepackage[caption=false]{subfig}

\usepackage{bbold}

\usepackage[sort,compress]{cleveref}
\crefname{section}{Sec.\!}{Secs.\!}
\crefname{equation}{Eq.\!}{Eqs.\!}
\crefname{figure}{Fig.\!}{Figs.\!}
\crefname{table}{Tab.\!}{Tabs.\!}
\crefname{appendix}{App.\!}{Apps.\!}

\usepackage{upgreek}
\usepackage[acronym]{glossaries-extra}
\glssetcategoryattribute{acronym}{nohyperfirst}{true}
\setabbreviationstyle[acronym]{long-short}

\newacronym{rg}{RG}{Renormalization Group}
\newacronym{frg}{FRG}{Functional Renormalization Group}
\newacronym{uv}{UV}{ultraviolet}
\newacronym{ir}{IR}{infrared}
\newacronym{lpa}{LPA}{local potential approximation}
\newacronym{mc}{MC}{Monte Carlo}
\newacronym{cb}{CB}{Conformal Bootstrap}
\newacronym{qcd}{QCD}{Quantum Chromodynamics}
\newacronym{kt}{KT}{Kurganov-Tadmor}
\newacronym{pde}{PDE}{partial differential equation}
\newacronym{ode}{ODE}{ordinary differential equation}

\usepackage[normalem]{ulem}

\newcommand{\ir}[0]{{\mathrm{IR}}}
\newcommand{\uv}[0]{{\mathrm{UV}}}

\newcommand{\vdistance}{\vphantom{\bigg(\bigg)}}

\begin{document}
\title{Reanalysis of critical exponents for the \texorpdfstring{$O(N)$}{O(N)} model via a hydrodynamic approach to the Functional Renormalization Group}

\author{Fabrizio Murgana}
    \email{murgana@itp.uni-frankfurt.de}
    \affiliation{
        Department of Physics and Astronomy, University of Catania,\\
        Via S.\  Sofia 64, I-95125 Catania, Italy
    }
    \affiliation{
        INFN-Sezione di Catania,\\
        Via S.\ Sofia 64, I-95123 Catania, Italy
    }
	\affiliation{
		Institut f\"ur Theoretische Physik, Goethe-Universit\"at,\\
		Max-von-Laue-Stra{\ss}e 1, D-60438 Frankfurt am Main, Germany
	}

\author{Adrian Koenigstein}
	\email{adrian.koenigstein@uni-jena.de}
	\affiliation{
		Institut f\"ur Theoretische Physik, Goethe-Universit\"at,\\
		Max-von-Laue-Stra{\ss}e 1, D-60438 Frankfurt am Main, Germany
	}
    \affiliation{
        Theoretisch-Physikalisches Institut, Friedrich-Schiller-Universit\"at Jena,\\
        Max-Wien-Platz 1, D-07743 Jena, Germany
    }

\author{Dirk H.\ Rischke}
    \email{drischke@itp.uni-frankfurt.de}
	\affiliation{
		Institut f\"ur Theoretische Physik, Goethe-Universit\"at,\\
		Max-von-Laue-Stra{\ss}e 1, D-60438 Frankfurt am Main, Germany
	}
	\affiliation{
		Helmholtz Research Academy Hesse for FAIR, Campus Riedberg,\\
		Max-von-Laue-Stra{\ss}e 12, D-60438 Frankfurt am Main, Germany
	}

\begin{abstract}
    We compute the critical exponents of the $O(N)$ model within the Functional Renormalization Group (FRG) approach.
    We use recent advances which are based on the observation that the FRG flow equation can be put into the form of an advection-diffusion equation.
    This allows to employ well-tested hydrodynamical algorithms for its solution.
    In this study we work in the local potential approximation (LPA) for the effective average action and put special emphasis on estimating the various sources of errors.
    Our results complement previous results for the critical exponents obtained within the FRG approach in LPA. Despite the limitations imposed by restricting the discussion to the LPA, the results compare favorably with those obtained via other methods.
\end{abstract}

\maketitle
\section{Introduction}

    The understanding of the behavior of systems where microscopic degrees of freedom 
    are strongly interacting is the main goal of many areas of physics, ranging from condensed matter \cite{Durrant:00,BCS:57} to elementary-particle theories  \cite{Schaefer:10, Yokota:16}, and extending even to quantum gravity \cite{Reuter:98, Percacci:15}.
    However, first-principle calculations for these systems are often very difficult and demand the use of powerful tools.
    Calculations are in particular challenging when a system undergoes a phase transition, since new degrees of freedom may arise or become relevant.
    In this case the underlying theory must consistently relate the two phases and thus describe the transition from one set of degrees of freedom to the other.
    For second-order transitions, the behavior of a system at all length scales is determined by a finite set of so-called critical exponents.
    
    A widely exploited theoretical method to treat this problem is the perturbative \gls{rg} introduced by Callan, Symanzik, and others.
    This technique describes systems where many interacting degrees of freedom are present which interact via small effective couplings \cite{Pelissetto:2000ek}.
    Unfortunately, this approach cannot be used in systems where such a small coupling, or in general a small parameter in which one can perform perturbative calculations, simply does not exist or is hard to identify.
    Furthermore, the perturbative series does not converge in general and one has to exploit resummation techniques or other methods.
    Another approach to describe phase transitions are computer simulations employing \gls{mc} methods, which have been successfully used to obtain high-precision estimates of the critical exponents, see, e.g., Refs.~\cite{Hasenbusch:2010hkh,Campostrini:2002ky}.
    However, one of the major issues with these methods is the extremely large amount of computer time needed to obtain a reliable infinite-volume and continuum limit.
    Yet another method which can address critical phenomena is the conformal-field theory approach \cite{Cardy:1992tq,Poland:19,Bissi:22}, which for example leads to a high precision for the critical exponents of the Ising model, originally in $2D$ and later in $3D$ thanks to the \gls{cb} technique.

    In order to overcome the difficulties of the perturbative \gls{rg} method and numerical simulations, a different, but similarly powerful approach can be used: the \gls{frg} \cite{Pawlowski:2005xe,Gies:12,Kopietz:2010zz,Dupuis:2020fhh,Berges:02}, also referred to as Exact \gls{rg} or Non-perturbative \gls{rg}, which is ultimately based on seminal ideas by Wilson \cite{Wilson:1971dh,Wilson:1971bg,WILSON:75,Wilson:1974mb,Wilson:1979qg} and others, see, e.g., Refs.~\cite{Polchinski:1983gv,Hasenfratz:1985dm}.
    The central object of the \gls{frg} approach is a flow equation, which describes the evolution of correlation functions or, equivalently, their generating functional under the influence of fluctuations.
    It connects a well-defined initial quantity, e.g., the microscopic \gls{uv} action, in an exact manner with the desired full \gls{ir} effective action, where all fluctuations are integrated out.
    Hence, solving the flow equation corresponds to solving the full theory and is therefore equivalent to a direct computation of the generating functional.
    Thanks to the fact that it is non-perturbative and connects degrees of freedom at different scales, the \gls{frg} approach is well suited to address the issue of describing systems approaching criticality and phase transitions, and thus a tool for the computation of critical exponents, cf.\ Refs.~\cite{Braun:2010tt,Janssen:2012pq,Litim:2002cf,Borchardt:2015rxa,Borchardt:2016kco,DePolsi:20,Bohr:01,Chlebicki:2020pvo,Balog:2019rrg}.
    
    Even though the \gls{frg} has a solid theoretical foundation, it is very hard to find analytical solutions to the flow equation.
    For specific problems, it can be converted into an infinite number of coupled \glspl{pde} and/or \glspl{ode}, which, if suitably truncated, can be subjected to a numerical treatment in order to obtain a solution.
    In fact, as a truncation scheme one often employs a derivative expansion of the effective action and solves separate \gls{frg} flow equations for the coefficients of this expansion.
    The leading-order truncation in this expansion only accounts for the flow of the local effective potential, and is thus called \gls{lpa}.

    In this work, we briefly introduce the \gls{frg} approach and apply it to the $O(N)$ model, for which we compute the critical exponents in the \gls{lpa}.
    The critical exponents are just one example of physical quantities that can be computed by the \gls{frg} method.
    However, they are relevant since they can provide a good benchmark to test the convergence of this method, because other techniques, such as \gls{cb} and \gls{mc}, yield a very precise determination of these quantities.

    The critical exponents of the $O(N)$ model have already been studied within the \gls{frg} approach some time ago, see, e.g., Refs.\ \cite{Litim:2002cf,DePolsi:20,Bohr:01,Balog:2019rrg,DePolsi:2021cmi}.
    The reason why we decided to repeat such an investigation is that recently a novel method to solve the \gls{frg} flow equations has been proposed \cite{Grossi:2019urj}.
    This method relies on the observation that the \gls{frg} flow equation for the effective potential can be cast into the form of a hydrodynamic advection-diffusion equation.
    This suggests to use numerical techniques well-known from hydrodynamics to solve that equation, see also Refs.~\cite{Grossi:2021ksl,Koenigstein:2021syz,Koenigstein:2021rxj,Steil:2021cbu,Stoll:2021ori,Ihssen:2022xjv,Ihssen:2022xkr} as well as Refs.~\cite{Zamolodchikov:1986gt,Zumbach:1994vg,Tetradis:1995br,Litim:1995ex,Rosten:2010vm,Aoki:2014ola,Aoki:2017rjl} for some early developments.
    In particular, hydrodynamic conservation laws in general allow for the formation of discontinuities or, more general, non-analyticities in the solution.
    Therefore, the applied numerical scheme has to be able to handle discontinuities in an appropriate manner.
    While Ref.\ \cite{Grossi:2019urj} used a discontinuous Galerkin method to solve the \gls{frg} flow equation, here we will exploit a well-established finite-volume central scheme, the \gls{kt} algorithm, which is designed to have a high-order accuracy, preserving stability and introducing negligible dissipation, while allowing to treat discontinuities in the solution (see Ref.\ \cite{Kurganov:00} for more details).
    Applying this method, we perform a more systematic study of various sources of errors in determining the critical exponents of the $O(N)$ model as compared to previous attempts using the \gls{frg} approach in the \gls{lpa} (without working with the fixed-point equations directly).\footnote{
        As already mentioned, there are more advanced truncations available for high-precision calculations of the critical exponents of $O(N)$ models via the \gls{frg} \cite{DePolsi:20,Bohr:01,Balog:2019rrg,DePolsi:2021cmi}, and we do not attempt to compete with the respective results.
        However, since some of these works combine the use of fixed-point equations with the approach employed in this work, we believe that our results are still of sufficient interest to be made public.}
    
    This work is organized as follows.
    In \cref{sec:II} we give a brief introduction to the \gls{frg} approach and discuss the \gls{frg} flow equation.
    In \cref{sec:III} we derive the \gls{frg} flow equation for the $O(N)$ model in the \gls{lpa} and rewrite it in the form of an advection-diffusion equation in order to solve it with the \gls{kt} scheme.
    Numerical results for the critical exponents are presented in \cref{sec:IV}.
    Finally, we conclude this work in \cref{sec:VI} with a summary and an outlook.
    A detailed discussion of error estimates is delegated to \cref{sec:V}.
    
\section{Functional Renormalization Group Approach}
\label{sec:II}

    In this section, we briefly recapitulate the \gls{frg} approach in the formulation given by Wetterich et al.\ \cite{Wetterich:1992yh,Ellwanger:1993mw,Morris:94,Reuter:1993kw}.
    For further details, see Refs.\ \cite{Pawlowski:2005xe,Gies:12,Kopietz:2010zz,Dupuis:2020fhh,Berges:02}.
    The central object of this approach is the effective action $\Gamma [ \Phi ]$, which is the generating functional of 1PI vertex functions.
    In order to compute $\Gamma [ \Phi ]$, one introduces the so-called \textit{effective average action} $\bar{\Gamma}_k [ \Phi ]$.
    This quantity depends on the parameter $k$, which is a coarse-graining scale with physical dimension of a momentum.
    The \gls{frg} flow equation describes the evolution of $\bar{\Gamma}_k [ \Phi ]$ as $k$ runs from the \gls{uv}, $k \to \infty$, to the \gls{ir}, $k \to 0$.
    The effective average action interpolates between the bare classical action $S_{\textrm{bare}} [ \Phi ]$ in the \gls{uv} and the full quantum effective action $\Gamma [ \Phi ]$ in the \gls{ir}, i.e.,
        \begin{align}\label{eq:1}
        	&  \bar{\Gamma}_{k \to \infty} [ \Phi ] = S_{\textrm{bare}} [ \Phi ] \, , && \bar{\Gamma}_{k \to 0} [ \Phi ] = \Gamma [ \Phi ] \, .
        \end{align}
    In practice, one may not always be able to send $k \to \infty$.
    Therefore, one usually introduces a \gls{uv} scale $\Lambda$ as initial scale for the \gls{frg} flow equation, which is chosen to be sufficiently large, i.e., much larger than any other physical scale of the theory, and assumes that the bare classical action describes the underlying theory at this scale.
    However, using a finite cutoff is an approximation and one has to ensure that the results are independent of the choice of $\Lambda$, e.g., by respecting \gls{rg} consistency \cite{Braun:2018svj}.

    The equation that describes the \gls{rg}-scale evolution of $\bar{\Gamma}_k [\Phi ]$, i.e., how the effective average action varies as one integrates out fluctuations with increasingly smaller momenta, is the so-called \textit{Wetterich equation}, or \textit{Exact Renormalization Group flow equation}, or simply \textit{\gls{frg} flow equation} \cite{WETTERICH:91,Wetterich:1992yh,Morris:1993qb},
        \begin{align}\label{eq:3}
        	\partial_k \bar{\Gamma}_k [ \Phi ] = \mathrm{Tr} \Big[ \big( \tfrac{1}{2} \,  \partial_k R_k \big) \big( \bar{\Gamma}^{(2)}_k [ \Phi ] + R_k \big)^{-1} \Big] =
            \begin{gathered}
                \vspace{-0.2cm}
                \includegraphics[scale=1.0]{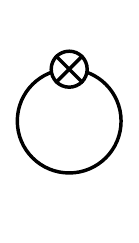}
            \end{gathered} \, ,
        \end{align}
    where the trace indicates an integral over momenta and a sum over all internal degrees of freedom.
    The equation has one-loop structure, which is indicated by the Feynman-diagram representation on the right-hand side.
    Here, the black line represents the full propagator $G_k [ \Phi ] = ( \bar{\Gamma}^{(2)}_k [ \Phi ] + R_k )^{-1}$ and the crossed circle stands for $\frac{1}{2} \, \partial_k R_k$, where $R_k$ in \cref{eq:3} is the so-called \textit{regulator}.
    The latter must be chosen such that $\bar{\Gamma}_k [ \Phi ]$ smoothly interpolates between the bare classical action $S_{\textrm{bare}} [ \Phi ]$ at $k = \Lambda$ and the full effective action $\Gamma [ \Phi ]$ at $k = 0$.
    This requirement imposes certain conditions on $R_k$.
    In particular, (i) it should act like a mass term for momenta smaller than $k$, regulating the full propagator $G_k [ \Phi ] = ( \bar{\Gamma}^{(2)}_k [ \Phi ] + R_k )^{-1}$ in the \gls{ir}, (ii) it has to vanish when $k \to 0$, and (iii) it has to diverge when $k \to \infty$, such that the functional integral defining $\bar{\Gamma}_k [ \Phi ]$ is dominated by the bare classical action.
    The choice of the specific shape of the regulator is a delicate issue and can be optimized depending on the problem, see, e.g., Refs.~\cite{Litim:2001up,Pawlowski:2015mlf,Braun:2020bhy,Braun:2022mgx} for details.
    
    Despite its deceptively simple form, the Wetterich equation is a functional integro-differential equation, and as such cannot be solved exactly for an arbitrary $\bar{\Gamma}_k [ \Phi ]$.
    Thus, it is clear that some approximation has to be made.
    Two common approximation schemes used in the literature are the vertex expansion \cite{Morris:94,Bergerhoff:98} and the derivative expansion \cite{Berges:02,Berges:95}.
    The former expands $\bar{\Gamma}_k [ \Phi ]$ in powers of the fields $\Phi$, where the vertex functions $\bar{\Gamma}_k^{(n)} ( x_1, \ldots, x_n ) = \delta^n \bar{\Gamma}_k [ \Phi ]/ \delta \Phi ( x_1 ) \cdots \delta \Phi ( x_n ) \big|_{\Phi = \Phi_0}$ of the theory are the expansion coefficients.
    On the other hand, in the derivative expansion one expands $\bar{\Gamma}_k [ \Phi ]$ in terms of all composite operators constructed from space-time derivatives (in principle of arbitrary order) of the fields.
    These composite operators must be compatible with the symmetries of the theory.
    In both cases, one obtains an infinite system of coupled integro-differential equations which needs to be truncated, either at a certain order of the vertex functions or at a certain order of space-time derivatives of the fields.
    In this work we will focus on the latter, since the vertex expansion assumes regularity of $\bar{\Gamma}_k [ \Phi ]$, which is violated near phase transitions, where the effective action develops discontinuities or points of non-analyticity during the \gls{frg} flow \cite{Grossi:2019urj,Grossi:2021ksl,Stoll:2021ori}.
    Since in this paper we are interested in critical exponents of second-order phase transitions, we have to employ a method which allows to treat discontinuities and non-analyticities in the \gls{frg} flow of the effective action.
    The derivative expansion fulfills this requirement, even in the most simple truncation, the aforementioned \glsfirst{lpa}, which will be discussed in the next section.

\section{Application: \texorpdfstring{$O(N)$}{O(N)} model in LPA} 
\label{sec:III}

    In this section we  will briefly introduce the $O(N)$ model and then apply the \gls{frg} approach in \gls{lpa} to derive the respective flow equation.

\subsection{The \texorpdfstring{$O(N)$}{O(N)} model}

    Let us consider the $O(N)$ model in $d$-dimensional Euclidean space-time.
    The model describes the dynamics of $N$ scalar fields $\phi_a (x)$, where $a = 1, \ldots , N$, with the bare action
        \begin{align}\label{eq:4}
        	S_{\textrm{bare}} [ \vec{\phi} \,] = \int d^d x \big[ ( \partial_\mu \phi_a )^2 + V ( \rho ) \big] \, ,
        \end{align}
    where $V ( \rho )$ is the potential parametrizing the form the $N$ scalar fields interact with each other.
    The $O(N)$ symmetry requires that $V ( \rho )$ is solely a function of the variable
        \begin{align} \label{eq:o_n_invariant}
            \rho = \tfrac{1}{2} \, \phi_a \, \phi_a \, ,
        \end{align}
    where a sum over $a$ from $1$ to $N$ is implied.
    Despite its apparent simplicity, the $O(N)$ model can describe a large variety of physical systems at different energy scales.
    The reason for the wide range of applicability of such a simple model is the well-known universal behavior of physical systems close to criticality, when the microscopic degrees of freedom are not important and only the main features, like the symmetry of the model, have to be taken into account in the description of the physical behavior of the system.
    For this reason, the $O(N)$ model can be considered as a prototype model to understand relevant mechanisms that govern a phase transition.
    For example, for $N = 4$ the model can describe the chiral phase transition in \gls{qcd} with two quark flavors \cite{Meyer-Ortmanns:1996ioo}.
    For $N = 3$ it belongs to the universality class of the Heisenberg model,  describing a ferromagnetic phase transition \cite{Kleinert:89}.
    The $N = 2$ case can be used to describe  the XY-model \cite{Hostetler:21} and $N = 1$  belongs to  the Ising universality class \cite{Barry:73}.
    
\subsection{Derivation of the flow equation in LPA}

    In order to proceed with the quantitative study of the $O(N)$ model, the first step will be to derive the flow equation that describes the $k$-evolution of the effective average action.
    As already mentioned in the previous section, we need to truncate the effective action, i.e., we need to choose an ansatz.
    We use the derivative expansion, expanding the effective action in terms of powers of gradients of the field.
    In particular we consider the lowest order of the expansion, the so-called \gls{lpa}, where the only space-time derivatives of the fields appear in the kinetic term and the effective potential $V ( \rho )$ is solely a function of the fields but not of their space-time derivatives.
    
    The advantage of the \gls{lpa} is that it leads to a simple expression for the \gls{frg} flow equation, while still capturing many non-trivial features of the theory.
    Furthermore, this approximation becomes exact in the limit $N \to \infty$ \cite{Attanasio:97}.
    In \gls{lpa} the effective average action reads
         \begin{align}\label{eq:5}
            \bar{\Gamma}_k [ \vec{\varphi}\, ] = \int \mathrm{d}^d x \big[ \tfrac{1}{2} \, ( \partial_\mu \varphi_a )^2 + V_k ( \varrho ) \big] \, ,
        \end{align}
    where $V_k ( \varrho )$ is the effective potential which depends on the \gls{frg} scale parameter $k$ and the $O(N)$-invariant $\varrho
    = \tfrac{1}{2} \varphi_a \varphi_a$, i.e., the quantity $\rho$ defined in \cref{eq:o_n_invariant}, evaluated for the mean fields $\varphi_a$.
    The next order in the derivative expansion would be the so-called \gls{lpa}${}^\prime$, which considers non-trivial wave-function renormalization corrections, i.e., $( \partial_\mu \varphi_a )^2 \to Z_k \, ( \partial_\mu \varphi_a )^2$, with $Z_k$ being a non-trivial function of the \gls{frg} flow parameter $k$.
    For even better truncations in similar models, see, e.g., Refs.~\cite{Canet:2002gs,Eser:2018jqo,Divotgey:2019xea,Cichutek:2020bli}.
    
    Once the ansatz \labelcref{eq:5} for the effective action is specified, the Wetterich equation \labelcref{eq:3} turns into a \gls{pde} for the effective potential $V_k ( \varrho )$.
    For a concrete result, we have to choose a regulator which satisfies the conditions we described in the previous section.
    It is possible to prove  \cite{Litim:2001up} that, in the case of the \gls{lpa}, an optimal choice for the regulator in terms of stability of the \gls{frg} flow equation is the so-called Litim regulator
    \begin{align}\label{eq:6}
        &   R_k(q, p) = ( 2 \uppi )^d \, \delta^{(d)} ( q + p ) \, p^2 \, r_k ( p ) \, ,    &&  r_k ( p ) = \left( \frac{k^2}{p^2} - 1 \right) \, \Theta \left( \frac{k^2}{p^2} - 1 \right) \, ,
    \end{align}
   where $r_k ( p )$ is the corresponding regulator shape function.
   The next step is the computation of the regularized propagator, or equivalently the two-point vertex function.
   To this end, we choose a simple background field 
   \begin{align}\label{eq:SSB}
            &   \vec{\varphi} = ( 0, \cdots, 0, \sqrt{2 \varrho} ) \, , &&  \Longleftrightarrow &&  \varphi_a = \sqrt{2 \varrho} \, \delta_{a N} \, .
    \end{align}
    In this way we obtain for the two-point vertex function in momentum space 
        \begin{align}\label{eq:7}
            \bar{\Gamma}^{(2)}_{k, ab} ( q, p ; \varrho ) = ( 2 \uppi )^d \, \delta^{(d)} ( q + p ) \, \Big\{ \big[ q^2 + V'_k ( \varrho ) \big] \, \delta_{a b} + 2 \varrho V''_k ( \varrho ) \, \delta_{a N} \, \delta_{b N} \Big\} \, , 
        \end{align}
    where $V_k'( \varrho ) = \partial_\varrho V_k ( \varrho )$.
    Inserting \cref{eq:7} into \cref{eq:3} and performing the trace one obtains
        \begin{align}\label{eq:8}
            \partial_k V_k ( \varrho ) = \int \frac{\mathrm{d}^d q}{(2 \uppi)^d} \, q^2 \, \big[ \tfrac{1}{2} \, \partial_k r_k( q ) \big] \left\{ \frac{N - 1}{q^2 \, [ 1 + r_k ( q ) ] + V'_k ( \varrho )} + \frac{1}{q^2 \, [ 1 + r_k ( q ) ] + V'_k ( \varrho ) + 2 \varrho V''_k ( \varrho )} \right\} \, .
        \end{align}
    Inserting \cref{eq:6} into \cref{eq:8}, the integration becomes trivial and employing \cref{eq:2} we arrive at the following well-known flow equation for the effective potential,
        \begin{align}\label{eq:9}
            \partial_k V_k ( \varrho ) = A_d \, k^{d + 1} \, \left[ \frac{N - 1}{k^2 + V'_k ( \varrho )} + \frac{1}{k^2 + V'_k ( \varrho ) + 2 \varrho V''_k ( \varrho )} \right] \, ,
        \end{align}
    with
        \begin{align}
            &   A_d = \frac{\Omega_d}{d (2 \uppi)^d} \; , && \Omega_d = \frac{2 \uppi^{d/2}}{ \Gamma(d/2)} \; .
        \end{align}
    Here, $\Omega_d$ is the volume of the $d-1$ dimensional unit sphere and $\Gamma$ is  the Gamma function.

\subsection{FRG flow equation in terms of the field \texorpdfstring{$\sigma$}{sigma} and boundary conditions}
\label{sec:IIIC}

    Up to now the \gls{pde} \labelcref{eq:9} for the effective potential is written in terms of the $O(N)$-invariant $\varrho$.
    However, when solving the \gls{pde} with a numerical scheme that relies on a discretization in $\varrho$-direction, we face the following problem.
    Usually, such a numerical scheme requires a stencil of points in the vicinity of any given point in the domain where $V_k ( \varrho )$ is defined.
    In particular, this stencil is then also required at the boundary $\varrho = 0$, i.e., we would have to specify $V_k ( \varrho )$ for some negative value of $\varrho$, which does not exist.
    
    A solution is to reformulate the equations above in terms of the field expectation value $\sigma = \sqrt{2 \varrho}$, i.e., we consider $V_k ( \sigma )$, $\partial_\sigma V_k ( \sigma )$, and $\partial^2_{\sigma} V_k ( \sigma )$ instead of $V_k ( \varrho )$, $\partial_\varrho V_k ( \varrho )$, and $\partial^2_{\varrho} V_k ( \varrho )$, respectively.
    We then rewrite \cref{eq:9} as follows:
        \begin{align}\label{eq:10}
            \partial_k V_k ( \sigma ) = A_d \, k^{d + 1} \bigg[ \frac{N - 1}{k^2 + \frac{1}{\sigma} \partial_{\sigma} V_k ( \sigma )} + \frac{1}{k^2 + \partial^2_{\sigma} V_k ( \sigma )} \bigg] \, .
        \end{align}
    This equation will be the reference point for our further investigations of the \gls{frg} flow in \gls{lpa}.
    
    The boundary condition at $\sigma = 0$ (including points in its vicinity) is now specified exploiting the residual $\mathbb{Z}_2$ symmetry of the potential,
        \begin{align}\label{eq:11}
            V_k ( \sigma ) = V_k ( - \sigma ) \, ,
        \end{align}
    which translates into a $\mathbb{Z}_2$ antisymmetry for $\partial_\sigma V_k ( \sigma )$,
        \begin{align}\label{eq:12}
            \partial_\sigma	V_k ( \sigma ) = - \partial_\sigma V_k ( - \sigma ) \, ,
        \end{align}
    and again a $\mathbb{Z}_2$ symmetry for $\partial^2_{\sigma} V_k ( \sigma )$,
        \begin{align}\label{eq:12a}
            \partial^2_{\sigma} V_k ( \sigma ) = \partial^2_{\sigma} V_k ( - \sigma ) \, .
        \end{align}
    The \gls{pde} \labelcref{eq:10} cannot be solved numerically on an infinite domain $\sigma \in [0, \infty )$.
    Therefore, we have to choose a sufficiently large value of $\sigma$ for the upper boundary, say $\sigma_{\textrm{max}}$, and we also need to specify $V_k ( \sigma_{\textrm{max}} )$.

    In the \gls{uv}, at $k \to \infty$, the effective potential $V_k ( \sigma )$ is given by the potential term in the bare action $S_{\textrm{bare}}$, which is usually a polynomial in powers of $\sigma$.
    For a $\mathbb{Z}_2$-symmetric potential, these powers must be even.
    For an interacting theory, the smallest power of $\sigma$ must then be at least four.
    This means that for large $\sigma$ the right-hand side of the \gls{frg} flow equation \labelcref{eq:10} is at least of order $\sim \sigma^{-2}$.
    Choosing a sufficiently large $\sigma_{\textrm{max}}$ ensures that the right-hand side of \cref{eq:10} can be neglected, which means that the effective potential $V_k ( \sigma_{\textrm{max}} )$ does not change under the \gls{frg} flow and stays at its \gls{uv} value.
    For a more detailed discussion on the boundary conditions and how to implement them we refer to Ref.\ \cite{Koenigstein:2021syz}.

\subsection{Advection-diffusion formulation of the FRG flow equation}
    
    For the sake of convenience, let us introduce the ``\gls{rg}-time'' parameter $t$ via
        \begin{align}\label{eq:2}
            &  t=-\ln \left( \frac{k}{\Lambda} \right) \, ,  &&   \frac{\partial}{\partial t} = - k \, \frac{\partial}{\partial k} \; .
        \end{align}
    Note that we choose the opposite sign as in most of the standard \gls{frg} literature, because we deem it more natural for a time to flow from $t = 0$ (at $k=\Lambda$) to $ t = \infty$ (at $k=0$).

    We will now show that it is possible to cast \cref{eq:10} into the form of an advection-diffusion equation well-known from hydrodynamics.
    This is possible because \cref{eq:10} is independent of the potential itself and thus one can rewrite it in terms of a \gls{pde} for 
        \begin{align}\label{eq:13}
        	u ( t, \sigma ) = \partial_\sigma V ( t, \sigma ) \, .
        \end{align}
    Analogously we will denote 
        \begin{align}\label{eq:14}
        	u^\prime ( t, \sigma ) = \partial_\sigma u ( t, \sigma ) = \partial^2_{\sigma} V ( t, \sigma ) \, .
        \end{align}
    Here, we have defined $V ( t, \sigma ) = V_k ( \sigma )$, i.e., we replaced the $k$-dependence of $V_k ( \sigma )$ by an equivalent dependence on the variable $t$.
    We also considered $V ( t, \sigma )$ as a function of the two continuous variables $t$ and $\sigma$.
    In the interpretation of the \gls{frg} flow equation as an advection-diffusion equation, $t$ will retain its role as a time variable for the \gls{frg} flow, while $\sigma$ will assume the role of a spatial variable.
    The main advantage of considering the \gls{frg} flow equation as an advection-diffusion equation is that it allows us to exploit the powerful toolbox of numerical methods that have been developed to solve hydrodynamical equations.
    
    We now introduce the function
        \begin{align}\label{eq:15}
            f ( t, u, \sigma ) = ( N - 1 ) \, A_d \, \frac{( \Lambda e^{- t} )^{d + 2}}{( \Lambda e^{- t} )^2 + \frac{1}{\sigma} u ( t, \sigma )} \, ,
        \end{align}
    which corresponds to the non-linear advection flux in the hydrodynamical interpretation, as well as the function
        \begin{align}\label{eq:16}
            g ( t, u^\prime ) = - A_d \, \frac{( \Lambda e^{- t})^{d + 2}}{(\Lambda e^{- t} )^2 + u^\prime ( t, \sigma )} \, ,
        \end{align}
    which corresponds to a non-linear diffusion flux.
    Taking the derivative of \cref{eq:9} with respect to $\sigma$ we then obtain the equation for $u ( t, \sigma )$ as
        \begin{align}\label{eq:17}
            \partial_t u ( t, \sigma ) + \partial_\sigma f ( t, u, \sigma ) = \partial_\sigma g ( t, u^\prime ) \, ,
        \end{align}
    which has exactly the form of a non-linear  advection-diffusion equation for some fluid field $u$.
    For more details on the properties of this equation in the \gls{frg} framework we refer to Refs.~\cite{Grossi:2019urj,Grossi:2021ksl,Koenigstein:2021syz,Koenigstein:2021rxj,Steil:2021cbu,Stoll:2021ori,Ihssen:2022xjv,Ihssen:2022xkr}.

\section{Critical behavior}
\label{sec:IV}

    In order to study the critical behavior of the $O(N)$ model, we first have to set an initial condition for the \gls{frg} flow in the \gls{uv}, i.e., at $t=0$ or $k = \Lambda$.
    The potential in the \gls{uv} is chosen to have the well-known $\phi^4$-form,
        \begin{align}\label{eq:19}
    	   V_{k = \Lambda} ( \varrho ) = \frac{\lambda}{4} ( \varrho - \varrho_{0} |_{t = 0} )^2 \, ,
        \end{align}
    where $\varrho_0 |_{t = 0}$ is the minimum of the potential at $k = \Lambda$.
    For $\varrho_0 |_{t = 0} > 0$, the system is in the broken phase, while for $\varrho_0 |_{t = 0} = 0$ it is in the symmetric phase.
    
    In particular, in the case where $\varrho_0 |_{t = 0} > 0$, one component of the $N$-dimensional scalar field $\vec{\phi}$ develops a non-vanishing expectation value, see \cref{eq:SSB}, and the $O(N)$-symmetry group is spontaneously broken to $O(N-1)$.
    According to Goldstone's theorem, this leads to $N-1$ massless modes, the so-called Nambu-Goldstone bosons.
    The remaining component of the field $\vec{\varphi}$ is the radial sigma mode discussed above in \cref{sec:IIIC}, which develops a non-vanishing mass proportional to the minimum $\varrho_0 |_{t = 0}$ \cite{Coleman:74}.
    For example, considering the model at finite temperatures, the $O(N)$ symmetry is restored via a second-order phase transition and all modes become degenerate in mass when the temperature is increased above a critical value.
    
    It seems clear that, due to the non-perturbative nature of the phase transition, a simple perturbative approach fails to describe this phenomenon.
    On the other hand, the non-perturbative \gls{frg} approach allows us to go beyond any finite order in perturbation theory, and thus to correctly treat this kind of problem.
    Via this approach we will determine several critical exponents in the \gls{lpa}.
        \begin{figure}
            \centering
            \begin{minipage}{0.5\textwidth}
                \centering
                \includegraphics[width=\linewidth]{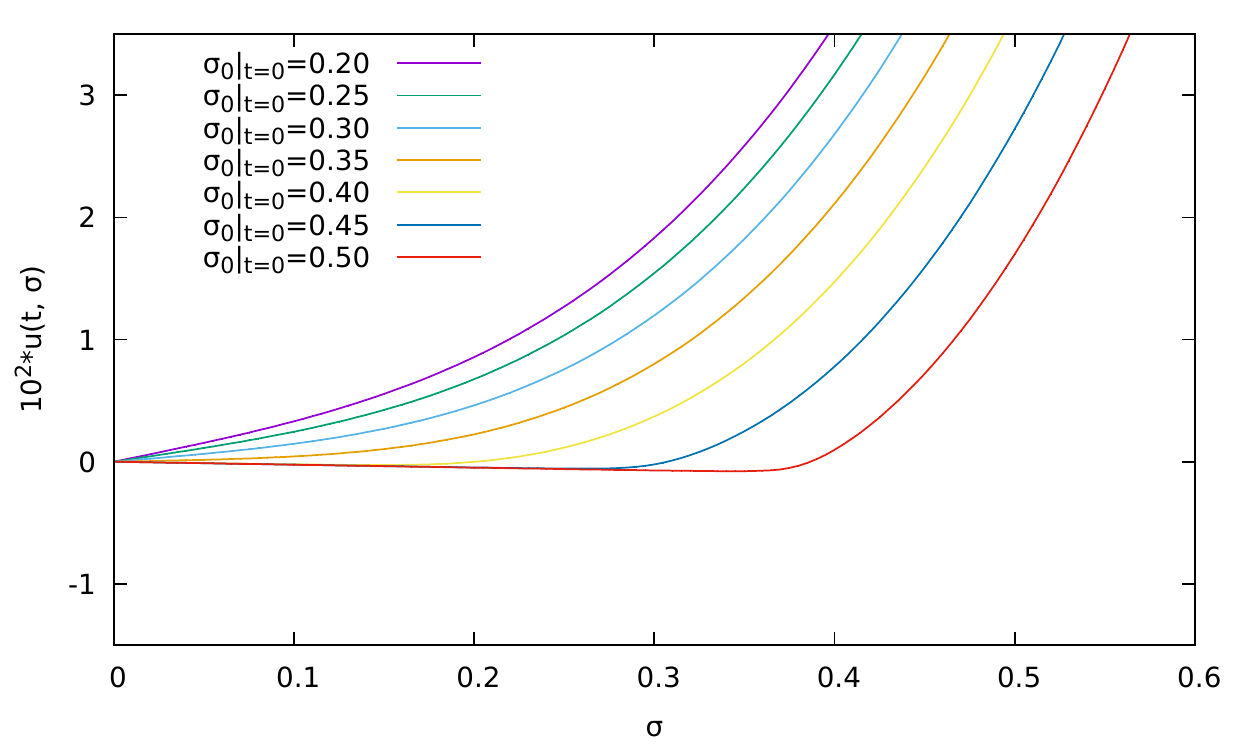}
            \end{minipage}%
            \begin{minipage}{0.5\textwidth}
                \centering
                \includegraphics[width=\linewidth]{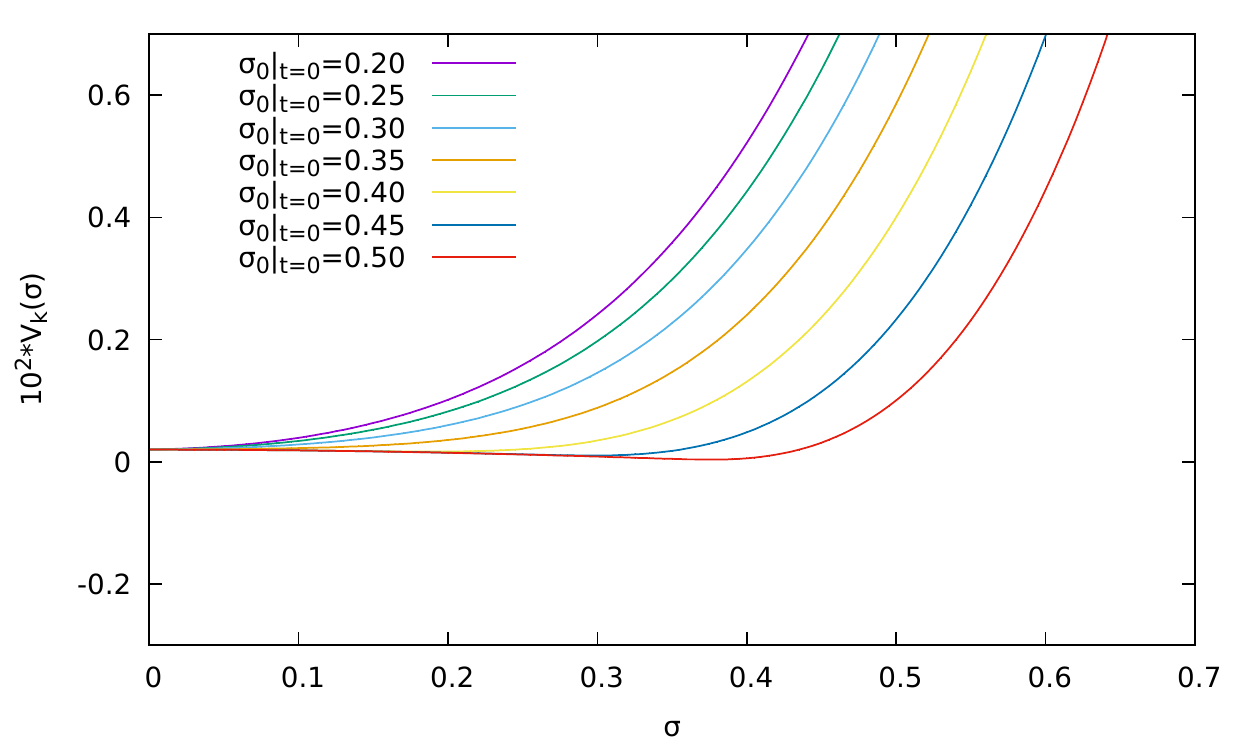}
            \end{minipage}
            \caption{The effective potential $V_k (\sigma)$ (right panel) and its derivative $u(t,\sigma)$ (left panel) in \gls{lpa} at \gls{rg} time $t = 3$, for different values of $\sigma_0 |_{t = 0} = \sqrt{2 \varrho_0|_{t = 0}}$ and for $N=3$. The UV cut-off scale is $\Lambda = 1.0$, the grid size is $[0, \sigma_{\text{max}}]  = [0,2.0]$, while the grid spacing is
            $\Delta \sigma = 0.005$ (corresponding to 400 grid points).}
            \label{fig:new1}
        \end{figure}

 \subsection{Critical scaling}

    In order to compute the critical exponents we  need the generalization of the flow equation \labelcref{eq:10} to finite temperature.
    However, we can circumvent this step by exploiting the well-known phenomenon of dimensional reduction \cite{Kocic:1995nf, Cavalcanti:19,Jungnickel:1998fj,Rajantie:1996np}: in the limit of high temperatures, or more precisely, when the temperature is much larger than the \gls{frg} flow parameter, $T \gg k$, the non-zero Matsubara modes with energy $\sim 2 \uppi n T$, $n \neq 0$, decouple from the evolution, leaving only a three-dimensional effective theory involving the zero Matsubara mode.
    From a different perspective, if the correlation length, which is the scale associated with the system, dominates over the inverse temperature, then it is not possible to resolve the compactified Euclidean time dimension, leading to a dimensionally reduced effective theory.
    Thus we do not need to use a finite-temperature flow equation in order to investigate the critical region of the phase transition.
    Instead, it is sufficient to employ the $d=3$-dimensional zero-temperature equation \labelcref{eq:10}.
    
    According to universality-class arguments \cite{Shang-keng:19, Goldenfeld:18, Cardy:96, Mussardo:10}, there are two relevant parameters that can be tuned in order to bring the system to the critical point.
    This can be immediately understood considering the ferromagnetic Ising model, which has a discrete $\mathbb{Z}_2 = O(1)$ symmetry, as an example: here the parameters that drive the system towards the phase transition are the temperature and the external magnetic field.
    Since we are considering the $O(N)$ model without external fields, only one relevant parameter is left in our case: the temperature.
    
    However, since we exploit the dimensional-reduction phenomenon and work in a zero-temperature field theory, the temperature does not explicitly enter the description and we need to identify a relevant variable which assumes its role.
    Obviously, such a variable decides whether the system ends up in the symmetric or broken phase when $t \to \infty$.
    This leads us to take the \gls{uv} minimum $\varrho_0 |_{t = 0}$ of the potential as the variable replacing the temperature.
    Indeed, if $\varrho_0 |_{t = 0}$ is larger than a critical value $\varrho^c_0 |_{t = 0}$, the system will end up in the broken phase in the IR, i.e., this is equivalent to working at temperatures $T < T_c$.
    Analogously, if $\varrho_0 |_{t = 0} < \varrho^c_0 |_{t = 0}$, the system will end up in the symmetric phase in the IR, which is equivalent to working at temperatures $T > T_c$.
    
    Due to the convexity of the effective potential, see for example Ref.\ \cite{Pelaez:16}, in the broken phase the potential $V_\ir ( \sigma )$ is flat for $\sigma \leq \sigma_0 |_\ir$, where $\sigma_0 |_\ir = \sqrt{2 \varrho_0 |_\ir} > 0$ is the minimum of the potential in the \gls{ir}, i.e., for $t \to \infty$.
    On the other hand, in the symmetric phase $\sigma_0|_\ir = 0$.
    This allows us to identify $\sigma_0 |_\ir$ as the \textit{order parameter} of the phase transition, since $\sigma_0 |_\ir > 0$ in the broken phase and $\sigma_0 |_\ir = 0$ in the symmetric one.
    
    This qualitative discussion is confirmed by solving the \gls{frg} flow equation, cf.\ \cref{fig:new1}, where we plot the effective potential $V_k ( \sigma )$ (right panel) and its derivative $u ( t, \sigma )$ (left panel) for different values of $\sigma_0 |_{t = 0}$ at the \gls{frg} time $t = 3$, which is sufficiently far in the \gls{ir} such that the flow does not change the minimum of the potential and the curvature mass at the minimum by an appreciable amount (for a detailed discussion of the numerical errors, see \cref{sec:V}).
    
    One observes that, for small values of $\sigma_0 |_{t = 0}$ the system  ends up in the symmetric phase in the \gls{ir} ($\sigma_0 |_\ir = 0$), while for larger values of  $\sigma_0 |_{t = 0}$ the symmetry remains broken in the \gls{ir} ($\sigma_0 |_\ir > 0$) and the potential exhibits a plateau.
    Since we are dealing with a second-order phase transition, we expect an \gls{ir} fixed point of the \gls{frg} flow, close to which the theory is scale-invariant.
    Thus the critical behavior has to be described by a solution of the \gls{frg} flow equation which is scale-independent for sufficiently small (large) values of $k$ ($t$).
    Since we only have one relevant variable ($\varrho_0 |_{t = 0}$), we are allowed to set $\lambda$, which is an irrelevant variable from the \gls{rg} perspective, to an arbitrary value.
    Here, we choose $\lambda = 0.5$.
    We then tune $\varrho_0 |_{t = 0}$ in order to find the so-called \textit{scaling solution} or \textit{critical trajectory}, i.e., a  solution of the \gls{frg} flow equation which becomes $t$-independent for sufficiently large values of $t$.
    
    In particular, the closer $\varrho_0 |_{t = 0}$ is to $\varrho^{c}_0 |_{t = 0}$, the closer the solution is to the critical trajectory when approaching the \gls{ir}.
    This implies that properly rescaled dimensionless quantities will exhibit a constant behavior at sufficiently large $t$.
    In particular, the dimensionless minimum
        \begin{equation}\label{eq:dimlessmin}
            \tilde{\varrho}_{0, k} = k^{2 - d} \varrho_{0, k}
        \end{equation}
    tends to a constant (fixed-point) value as $\varrho_0 |_{t = 0}$ approaches $\varrho^{c}_0 |_{t = 0}$.
    
            \begin{figure}[t]
            \centering
            \includegraphics[width=0.5\textwidth]{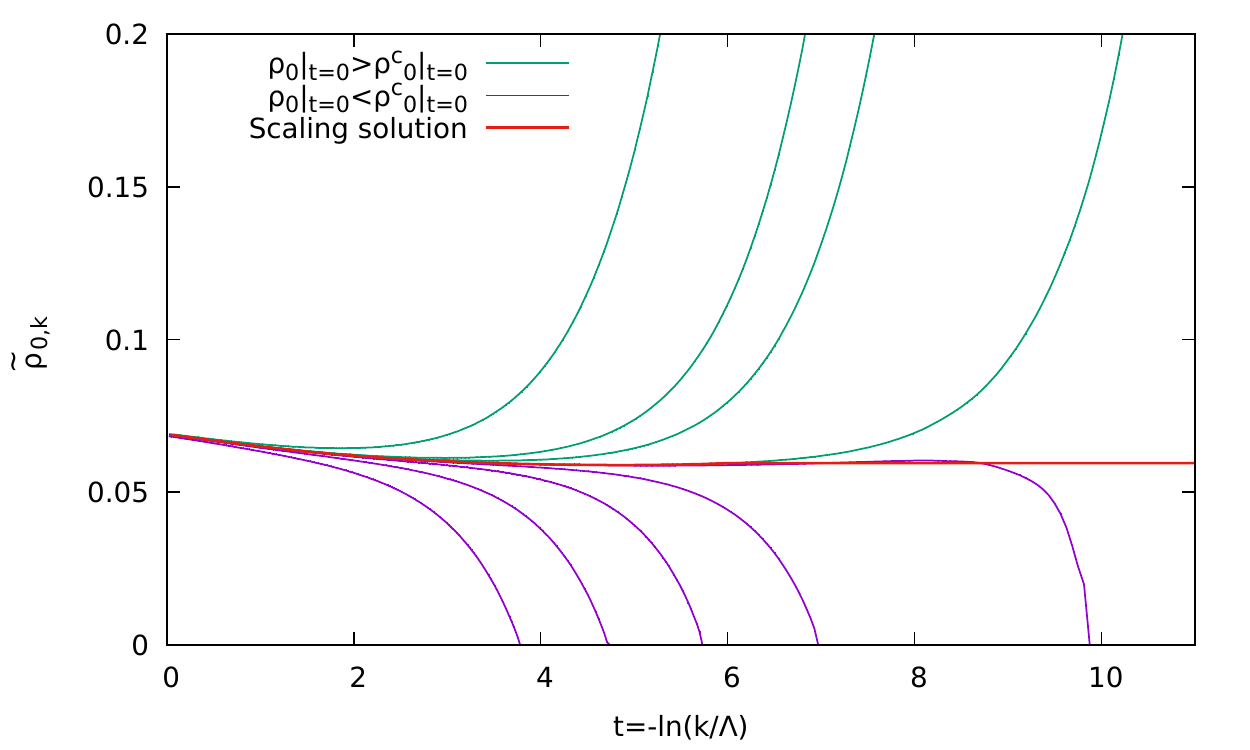}
            \caption{Scaling of $\tilde{\varrho}_{0,k}$ as a function of \gls{rg}-time $t$ in the \gls{lpa} for $N=3$ and $d=3$. The calculations are performed for $\Lambda = 1.0$, $\sigma_{\text{max}} = 2.0$, and grid spacing $\Delta \sigma = 0.0005$ (i.e., an order of magnitude smaller than for \cref{fig:new1}). 
            The green curves correspond to initial values $\varrho_0|_{t = 0} > \varrho^c_0|_{t = 0}$, indicating that the system is in the broken phase, while the purple curves correspond to initial values $\varrho_0|_{t = 0} < \varrho^c_0|_{t=0}$, i.e., the system is in the symmetric phase.
            The red horizontal line indicates the fixed-point value of the scaling solution in the \gls{ir}. }
            \label{scaling}
        \end{figure}
    This qualitative behavior is quantitatively confirmed by an explicit solution of the \gls{frg} flow equation, as shown in \cref{scaling}, cf.\ Ref.~\cite{Bohr:01}.
    Here we plot the evolution of the dimensionless minimum \labelcref{eq:dimlessmin} with $t$ for different initial values $\varrho_0 |_{t = 0}$.
    One observes that, during the \gls{frg} flow, $\tilde{\varrho}_{0, k}$ approaches the critical trajectory, the asymptotic fixed-point value of which is shown by the red horizontal line.
    However, eventually it deviates upwards (for initial values $\varrho_0 |_{t = 0} > \varrho^{c}_0 |_{t = 0}$) or downwards (for initial values $\varrho_0 |_{t = 0} < \varrho^{c}_0 |_{t=0}$).
    This can be easily explained by considering that, in $d = 3$ dimensions,
        \begin{itemize}
            \item   in the broken phase ($\varrho_0 |_{t = 0} > \varrho^c_0 |_{t = 0}$) the minimum $\varrho_{0, k}$ of the potential tends to a constant value $\varrho_0 |_\ir > 0$ when $k \to 0$, which means that $\tilde{\varrho}_{0, k} = \varrho_{0, k}/k \to +\infty$,
            \item   in the symmetric phase ($\varrho_0 |_{t = 0} < \varrho^c_0 |_{t = 0}$) the minimum $\varrho_{0, k}$ of the potential goes to $0$ already at a non-zero value of $k > 0$, which means that $\tilde{\varrho}_{0, k} = \varrho_{0, k}/k \to 0$.
        \end{itemize}
    The closer the initial value $\varrho_0|_{t=0}$ is to the critical value $\varrho^{c}_0|_{t=0}$, the larger is the value of $t$ where the deviation from the scaling solution occurs.
    The critical trajectory, ending in the fixed-point value as $t \to \infty$, can actually not be reached in practice, as it would require infinite numerical precision and infinite spatial resolution in the domain of the \gls{pde}.
    
    Of course not only $\tilde{\varrho}_{0, k}$ exhibits a scaling behavior, but all other appropriately rescaled dimensionless quantities first approach and then deviate from the critical trajectory at certain \gls{rg} times which depend on the initial value $\varrho_0 |_{t = 0}$.
    Here, we choose $\tilde{\varrho}_{0, k}$ since the minimum of the potential is the order parameter characterizing the phase transition.

\subsection{Critical exponents}

    It is a well-known fact \cite{Shang-keng:19, Goldenfeld:18, Cardy:96, Mussardo:10} that some physical quantities derived from the partition function or, equivalently, from the free energy of the system, may diverge when approaching a phase transition, i.e., for $\varrho_0 |_{t = 0} \to \varrho^{c}_0 |_{t = 0}$.
    In particular they exhibit a \textit{power-law behavior} characterized by so-called \textit{critical exponents}, which depend on the dimension $d$ and the symmetries of the system, and which define the \textit{universality class} of the theory.
    There exist six critical exponents: $\alpha$, $\beta$, $\gamma$, $\delta$, $\nu$, $\eta$.
    However, only two of them are independent since they are constrained by the following so-called \textit{scaling laws} or \textit{scaling relations}:
        \begin{align}
            &   \alpha = 2 - d \nu \, ,             &&  \textit{Josephson law} \, ,     \label{eq:20}   \vdistance
            \\
            &   \gamma = \nu ( 2 - \eta ) \, ,      &&  \textit{Fisher law} \, ,        \label{eq:21}   \vdistance
            \\
            &   \gamma = \beta ( \delta - 1 ) \, ,  &&  \textit{Widom law} \, ,         \label{eq:22}   \vdistance
            \\
            &   2 = \alpha + 2 \beta + \gamma \, ,  &&  \textit{Rushbrooke law} \, .    \label{eq:23}   \vdistance
        \end{align}

    In the following we will consider $\beta$ and $\nu$ as independent and discuss in more detail how to extract them from the \gls{frg} flow equation.
        \begin{figure}[t]
            \centering
            \includegraphics[width=0.5\textwidth]{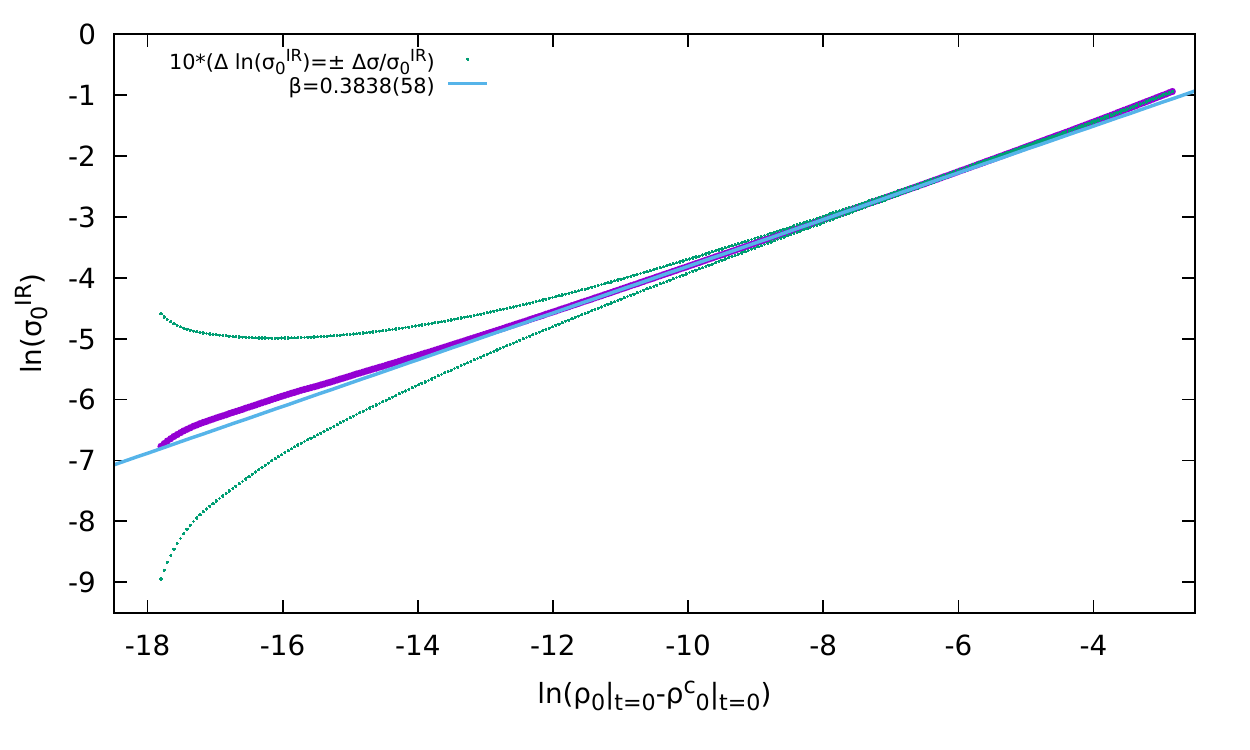}
            \caption{Double-logarithmic plot of \cref{eq:24} (violet dots) in the \gls{lpa} case for $N = 3$ and $d = 3$. The green dots mark the boundaries of the error band. The slope of the blue line gives the estimate for the critical exponent $\beta$. Other parameters of the calculation are the same as for \cref{scaling}.}
            \label{beta}
        \end{figure}
        \begin{enumerate}
            \item   \textit{Exponent for the order parameter: $\beta$}\\[0.1cm] Close to criticality, the order parameter of the phase transition $\sigma_0 |_\ir$ is described by the following behavior
                \begin{equation}\label{eq:24}
                    \begin{aligned}
                        &   \sigma_0 |_\ir = 0 \, ,    &&  \varrho_0 |_{t = 0} < \varrho^c_0 |_{t = 0} \, ,  \vdistance
                        \\
                        &   \sigma_0 |_\ir \sim ( \varrho_0 |_{t = 0} - \varrho^c_0 |_{t = 0} )^{\beta} \, , &&  \varrho_0|_{t = 0} > \varrho^c_0 |_{t = 0} \, .   \vdistance
                    \end{aligned}
                \end{equation}

        In \cref{beta}, we show $\ln(\sigma_0|_\ir)$ as a function of $\ln ( \varrho_0 |_{t = 0} - \varrho^c_0 |_{t = 0} )$ for $N = 3$ and $d = 3$ in the \gls{lpa}.
        The exponent $\beta$ is then read off from the slope of this function.
        
        One observes that, for small values of $\ln ( \varrho_0 |_{t = 0} - \varrho^c_0 |_{t = 0} )$, $\ln ( \sigma_0 |_\ir )$ deviates from the scaling behavior given by the second line of \cref{eq:24}.
        This deviation is caused by the finite numerical precision with which we can determine $\varrho^{c}_0 |_{t = 0}$.
        The closer the initial $\varrho_0 |_{t = 0}$ is to the critical value, the more sensitive one is to deviations from the actual critical value caused by the finite numerical precision, and thus one simply does not follow the critical behavior anymore.
        
        Note that, if we work in the finite-temperature theory, $\sigma_0 |_\ir$ would be a function of temperature and close to criticality would be proportional to $( T_c - T )^{\beta}$.
        Thus $\varrho_0^c |_{t = 0}$ determines the critical temperature in three dimensions.
        
        \item   \textit{Exponent for the correlation length: $\nu$}\\[0.1cm]
        For large spatio-temporal distances, the correlator exhibits an exponential decay,
            \begin{equation}\label{eq:25}
                G ( x - y ) = \ave{\phi ( x ) \, \phi ( y )} - \av{\phi ( x )} \, \av{\phi ( y )} \sim e^{- | x - y |/\xi} \quad \mbox{for} \quad | x - y | \gg \xi \, ,
            \end{equation}
        where $\xi$ is the so-called \textit{correlation length}.
        Close to  criticality it behaves as
            \begin{equation}\label{eq:26}
                \xi ( T ) \sim ( T - T_c )^{- \nu} \, ,
            \end{equation}
        which, in the dimensionally reduced theory, becomes
            \begin{equation}\label{eq:27}
                \xi ( \varrho_0 |_{t = 0} ) \sim \big| \varrho_0 |_{t = 0} - \varrho^c_0 |_{t = 0} \big|^{- \nu} \, .
            \end{equation}
            \begin{figure}
                \centering
                \includegraphics[width=0.5\textwidth]{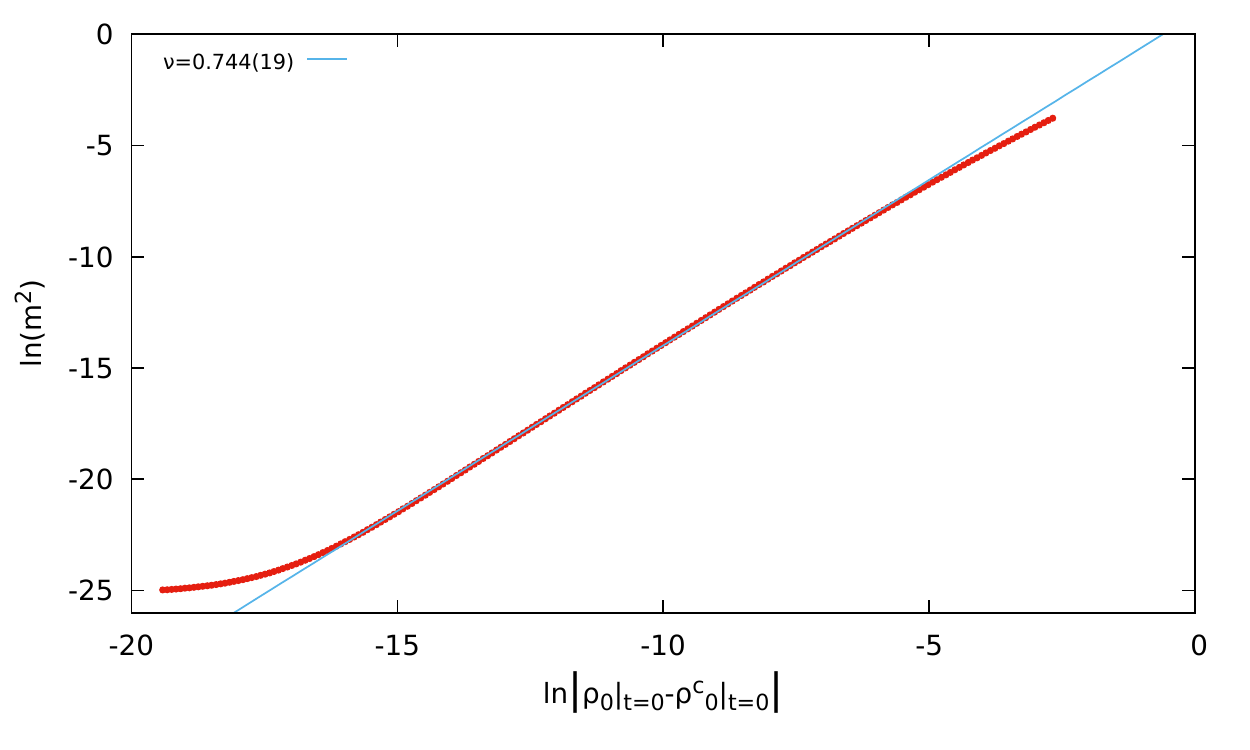}
                \caption{Double-logarithmic plot of \cref{eq:29} (red dots)  in the \gls{lpa} case for $N = 3$ and $d = 3$. The slope of the blue line gives twice the value of the critical exponent $\nu$. Other parameters of the calculation are the same as for \cref{scaling}.}
                \label{nu}
            \end{figure}
        In order to compute $\nu$, we exploit the fact that the correlation length is proportional to the inverse of the renormalized mass $m$, $\xi \sim m^{-1}$.
        Note that in the \gls{lpa} the pole, curvature, and screening masses are identical for zero-temperature calculations \cite{Helmboldt:14}.
        In the symmetric phase, the square of the renormalized mass is given by
            \begin{equation}\label{eq:28}
                m^2 = \lim_{t \to \infty} u^\prime ( t, \sigma = 0 ) = \lim_{t \to \infty} \partial^2_{\sigma} V ( t, \sigma = 0 ) \, .
            \end{equation}
        Thus we compute $m^2$ and then exploit
            \begin{equation}\label{eq:29}
                m^2 \sim \big| \varrho_0|_{t = 0} - \varrho^c_0 |_{t = 0} \big|^{2 \nu} \, .
            \end{equation}
        In this way we can obtain $\nu$ from the slope of $\ln m^2$ as a function of $\ln \big| \varrho_0 |_{t = 0} - \varrho^c_0 |_{t = 0} \big|$, see \cref{nu}.
        We note that one has to make sure to reach the symmetric phase before stopping the $t$-evolution, because only this ensures that $m^2 > 0$ at $\sigma = 0$.
        
        We remark that the deviation of $\ln m^2$ from the scaling solution at small values of $\ln \big| \varrho_0 |_{t = 0} - \varrho^c_0 |_{t = 0} \big|$ has the same origin as that observed in \cref{beta}.
        On the other hand, the deviation observed for large values of $\ln \big| \varrho_0|_{t = 0} - \varrho^c_0 |_{t = 0} \big|$ is due to the fact that one is simply too far away from the critical region, such that the scaling behavior of \cref{eq:25} does no longer apply.
        \end{enumerate}
    In \cref{tab:beta,tab:nu} we show our results for $\beta$ and $\nu$, respectively, obtained within the \gls{lpa} for $d=3$ and various values of $N$ in comparison to results obtained within similar as well as other frameworks.
    Results obtained within the same \gls{frg} framework but including a non-trivial \gls{rg}-scale dependent wave-function renormalization \cite{Bohr:01} are denoted as ``\gls{rg}' ''.
    Results obtained from \gls{mc} simulations \cite{Hasenbusch:2010hkh,Campostrini:2002ky} are listed under ``\gls{mc}'', those from perturbative \gls{rg} are denoted as ``PT'' \cite{Guida:1998bx}, while those from the $\varepsilon$-expansion at order $\varepsilon^6$ are shown under ``$\varepsilon$--exp.'' \cite{Kompaniets:2017yct}.
    Results from \gls{cb} are denoted as ``\gls{cb}'' \cite{Shimada:2015gda,Chester:2019ifh,Kos:2013tga}, and finally those from a derivative expansion up to fourth order as ``$\mbox{DE}_4$'' \cite{DePolsi:20}.
    We observe that the results obtained in this work are overall in good agreement with the ones obtained with other approaches.
    As compared to the results of \gls{rg}', which is from a technical perspective closest to our approach, we improve the precision by one order of magnitude and in addition provide an error estimate, see \cref{sec:V} for details.
    Although \gls{rg}' includes a non-trivial wave-function renormalization, while our approach does not, our results for the values of the critical exponents are not significantly worse than \gls{rg}'.
    Furthermore, we have checked that the values obtained for finite $N$ tend to converge to the ones in the large-$N$ limit, shown in \cref{N}, since in that case the \gls{lpa} becomes exact \cite{Morris:94}.
        \begin{table}
             \begin{tabular}{|c|c|c|c|c|c|c|c|}
                \hline
                $N$  & \gls{lpa} & \gls{rg}' & \gls{mc} & PT & $\varepsilon-$exp. & CB & $\mbox{DE}_4$
                \\
                \hline
                    & & & & & & &
                \\
                $1$  &  $0.3486(59)$ & $0.32$ & $0.32643(6)$ & $0.3258(10)$ & $0.32599(32)$ & $0.326419(2)$ & $0.3263(4)$
                \\
                    & & & & & & &
                \\
                \hline
                    & & & & & & &
                \\
                $2$  & $0.3659(45)$& $0.35$ & $0.34864(5)$ & $0.3470(11)$ & $0.3472(6)\hphantom{00}$ & $0.34872(5)\hphantom{0}$ & $0.3485(5)$
                \\
                    & & & & & & &
                \\
                \hline
                    & & & & & & &
                \\
                $3$  & $0.3838(58)$& $0.37$ & $0.3689(3)\hphantom{0}$ & $0.366(2)\hphantom{00}$ & $0.366(1)\hphantom{000}$ & $0.3697(12)\hphantom{0}$ & $0.3691(7)$
                \\
                    & & & & & & &
                \\
                \hline
                    & & & & & & &
                \\
                $4$  & $0.4046(34)$ & $0.40$ & $0.3873(4)\hphantom{0}$ & $0.3834(35)$ & $0.3834(18)\hphantom{0}$ & $0.3877(47)\hphantom{0}$ & $0.3874(6)$
                \\
                    & & & & & & &
                \\
                \hline
                    & & & & & & &
                \\
                $10$ & $0.4541(45)$ & $0.45$ & & & $0.4398(7)\hphantom{00}$ & $0.4523(2)\hphantom{00}$ & $0.4489(6)$
                \\
                    & & & & & & &
                \\
                \hline
            \end{tabular}
            \caption{\label{tab:beta}Order parameter exponent $\beta$.}
        \end{table}

        \begin{table}
            \begin{tabular}{|c|c|c|c|c|c|c|c|}
                \hline
                $N$  & \gls{lpa} & \gls{rg}' & \gls{mc} & PT & $\varepsilon-$exp. & CB & $\mbox{DE}_4$
                \\
                \hline
                    & & & & & & &
                \\
                $1$ & $0.634(8)\hphantom{00}$ & $0.64$ & $0.63002(10)$ & $0.6304(13)$ & $0.6292(5)\hphantom{0}$ & $0.629971(4)$ & $0.62989(25)$
                \\
                    & & & & & & &
                \\
                \hline
                    & & & & & & &
                \\
                $2$  & $0.7057(14)$ & $0.69$ & $0.67169(7)\hphantom{0}$ & $0.6703(15)$ & $0.6690(10)$ & $0.6718(1)\hphantom{00}$ & $0.6716(6)\hphantom{00}$
                \\
                    & & & & & & &
                \\
                \hline
                    & & & & & & &
                \\
                $3$  & $0.744(19)\hphantom{0}$ & $ 0.74$ & $0.7112(5)\hphantom{00}$ & $ 0.7073(35)$ & $ 0.7059(20)$ & $ 0.7120(23)\hphantom{0}$ & $0.7114(9)\hphantom{00}$
                \\
                    & & & & & & &
                \\
                \hline
                    & & & & & & &
                \\
                $4$  & $0.780(20)\hphantom{0}$ & $0.78$ & $0.7477(8)\hphantom{00}$ & $ 0.741(6)\hphantom{00}$ & $ 0.7397(35)$ & $ 0.7472(87)\hphantom{0}$ & $0.7478(9)\hphantom{00}$
                \\
                    & & & & & & &
                \\
                \hline
                    & & & & & &  &
                \\
                $10$ & $0.901(13)\hphantom{0}$ &  $0.91$ & & & $0.859(1)\hphantom{00}$ & $0.8842(3)\hphantom{00}$ & $0.8776(10)\hphantom{0}$ 
                \\
                    & & & & & & &
                \\
                \hline
            \end{tabular}
            \caption{\label{tab:nu}Correlation length exponent $\nu$.}
        \end{table}

        \begin{table}
            \begin{tabular}{|l|c|c|c|c|c|c|}
                \hline
                & $\beta$ &$\nu$ & $\eta$& $\delta$& $\alpha$& $\gamma$
                \\
                \hline
                Large $N$ & $0.5$ &$1.0$ & $0$& $5$& $-1.0$& $2.0$
                \\
                \hline
            \end{tabular}
            \caption{\label{N}Critical exponents in the large-$N$ case.}
        \end{table}

\section{Conclusion and Outlook}
\label{sec:VI}

    In this work we have applied the \glsfirst{frg} approach to compute the critical exponents of the $O(N)$ model.
    We exploited the recent realization that the \gls{frg} flow equation for the effective potential can be put into the form of an advection-diffusion equation \cite{Grossi:2019urj}, allowing us to employ widely used and well-tested hydrodynamic algorithms to solve it.
    We demonstrated the feasibility of this approach for the computation of the $O(N)$ critical exponents by solving the \gls{frg} flow equation in the \glsfirst{lpa}.
    As hydrodynamic algorithm, we used a finite-volume method, the so-called \gls{kt} scheme \cite{Kurganov:00}.
    The critical exponents of the $O(N)$ model have been studied previously within the \gls{frg} approach in \gls{lpa} and our discussion closely follows Ref.~\cite{Bohr:01}.
 However, here we demonstrated that the novel hydrodynamic approach to solve the \gls{frg} flow equations allows a better control of statistical errors.
    With this method, these errors can be more precisely determined and are of order $10^{-2}$ to $10^{-3}$. They are thus comparable to the errors of other well-established methods like lattice calculations, perturbation theory, the $\varepsilon$-expansion, or \gls{cb}.
    
    The range of applicability for the novel hydrodynamic method to solve \gls{frg} flow equations is very wide.
  Studies in the framework of the $O(N)$ model were already carried out including higher-order terms in gradients, such as wave-function renormalization factors \cite{DePolsi:20,Bohr:01,Balog:2019rrg,DePolsi:2021cmi}. Combining these high-precision studies with our developments is certainly a worthwhile future project.
    Furthermore, it is possible to study the system at finite temperature, without exploiting dimensional reduction.
    In particular, we expect that the non-zero Matsubara modes do not influence the exponents related to the singular part of free energy (the effective action), that is $\alpha, \gamma, \eta, \nu$, while in principle they could give contributions to the order parameter, thus modifying $\beta$, and to the critical isotherm, changing the value of $\delta$.
    
    Another possibility is to extend the model to include fermions.
    From a mathematical point of view, the \gls{frg} flow equations is still an advection-diffusion equation, but now with a source term  \cite{Wink:2020tnu,Grossi:2021ksl,Stoll:2021ori}, which can also be solved using the approach described above.

\section*{Acknowledgements}

    The authors gratefully acknowledge helpful discussions with L.\ Kiefer, L.\ Pannullo, O.\ Philipsen, M.~J.~Steil, and N.\ Wink.
    This work is supported by the Deutsche Forschungsgemeinschaft (DFG, German Research Foundation) through the Collaborative Research Center CRC-TR 211 ``Strong-interaction matter under extreme conditions'' - project number 315477589 - TRR 211 and by the State of Hesse within the Research Cluster ELEMENTS (Project ID 500/10.006).

\appendix

\section{Numerical precision and error estimates}
\label{sec:V}

    In order to provide some insight into the capabilities and limitations of our method, in this appendix we discuss the numerical precision of our results and give error estimates.
    There is of course a systematic error due to the use of the \gls{lpa} \cite{Litim:2002cf,DePolsi:20}, which can be estimated by improving the truncation beyond \gls{lpa} \cite{DePolsi:20,Bohr:01,Balog:2019rrg,DePolsi:2021cmi}.
    This, however, is beyond the scope of the current paper.
    Moreover, there are also fit errors which result from the way the critical exponents are determined.
    One of them is general in the sense that it is inherent to the extraction of both $\beta$ and $\nu$, but there are also some which are specific for the calculation of either $\nu$ or $\beta$.

\subsection{General fit error}
\begin{figure}
        \centering
        \begin{minipage}{0.5\textwidth}
            \centering
            \includegraphics[width=\linewidth]{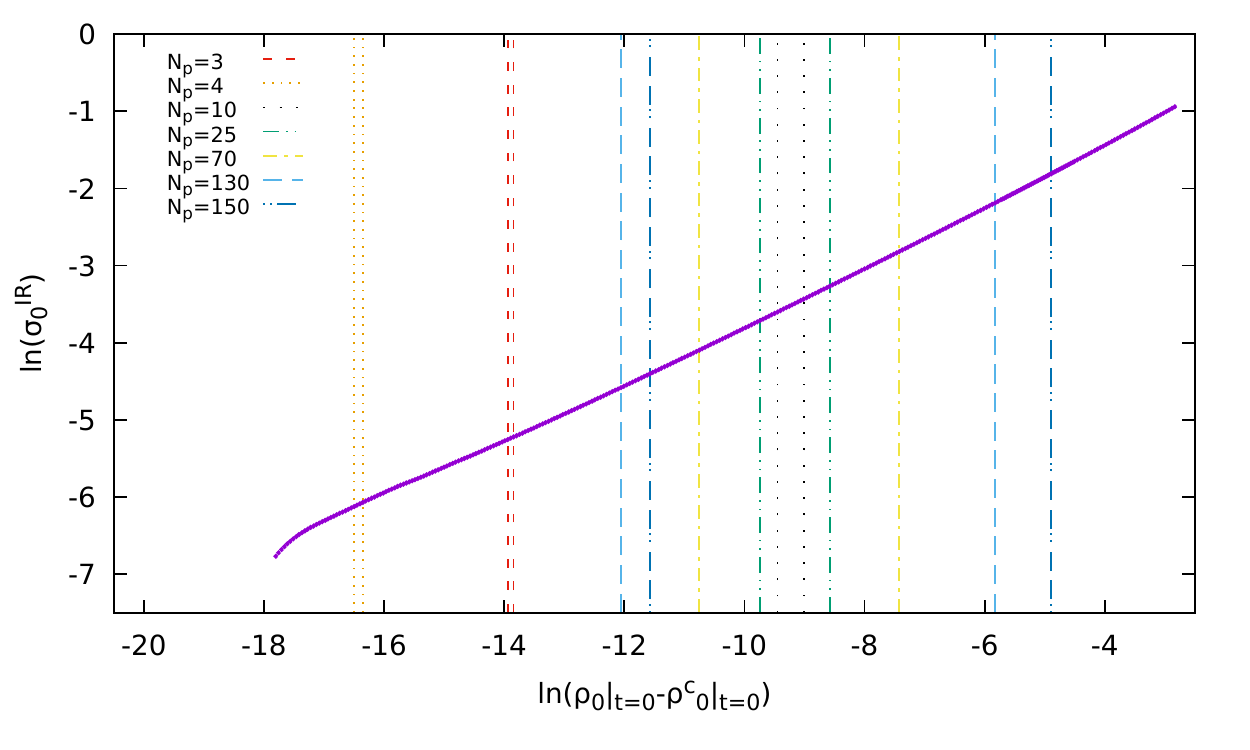}
        \end{minipage}%
        \begin{minipage}{0.5\textwidth}
            \centering
            \includegraphics[width=\linewidth]{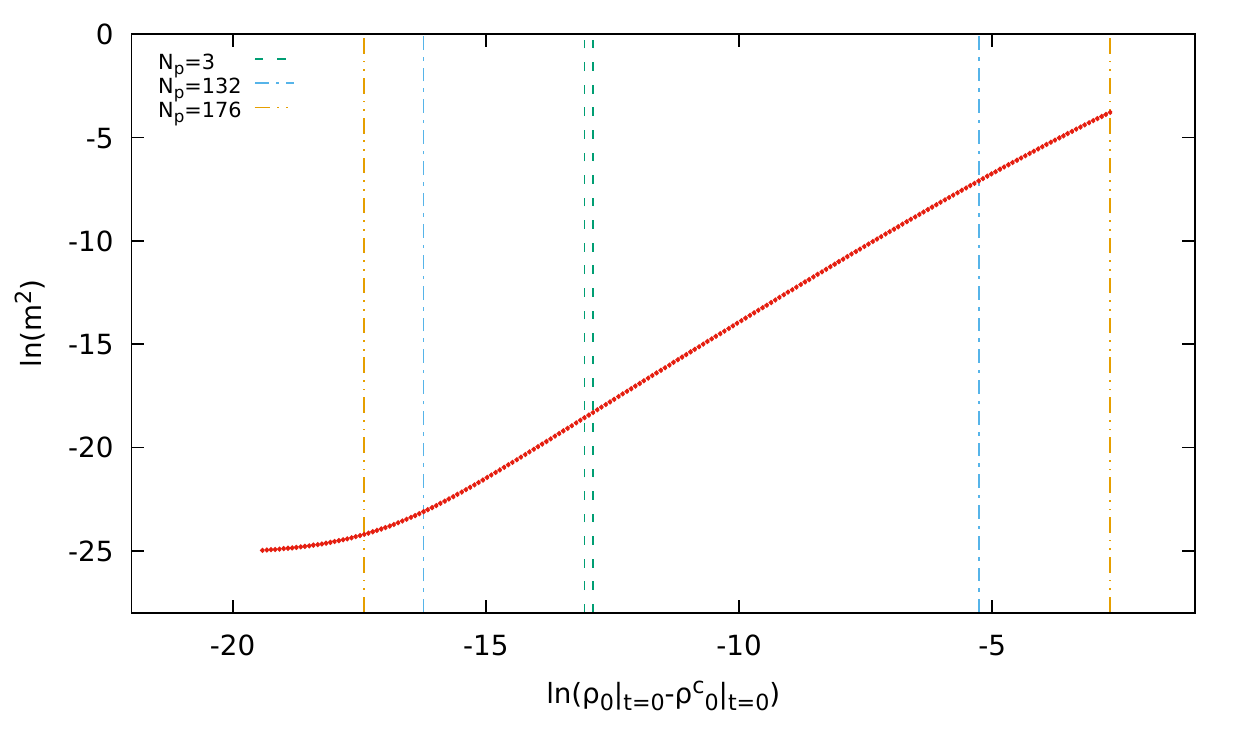}
        \end{minipage}
        \caption{Different fitting regions corresponding to different numbers of most aligned consecutive points, for the extraction of the critical exponents $\beta$ (left panel) and $\nu$ (right panel).
        Here, $N=3$ and $d=3$. Other parameters of the calculation are the same as for \cref{scaling}.}
        \label{al}
    \end{figure}

         \begin{figure}[t]
            \centering
            \includegraphics[width=0.5\textwidth]{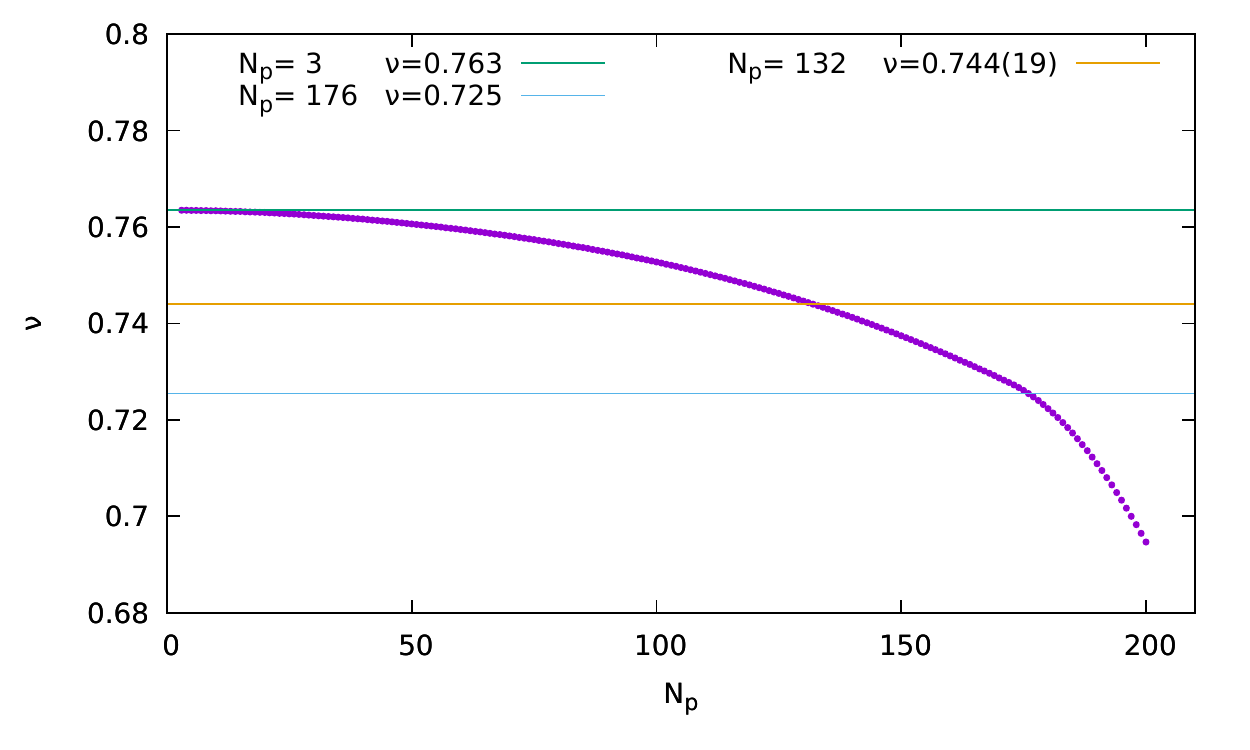}
            \caption{Critical exponent $\nu$ as a function of the number of the most aligned points $N_p$ taken into consideration in the fit, for the case $N = 3$ and $d = 3$. Other parameters of the calculation are the same as for \cref{scaling}.}
            \label{nup}
        \end{figure}

        \begin{figure}
        \centering
        \begin{minipage}{0.5\textwidth}
            \centering
            \includegraphics[width=\linewidth]{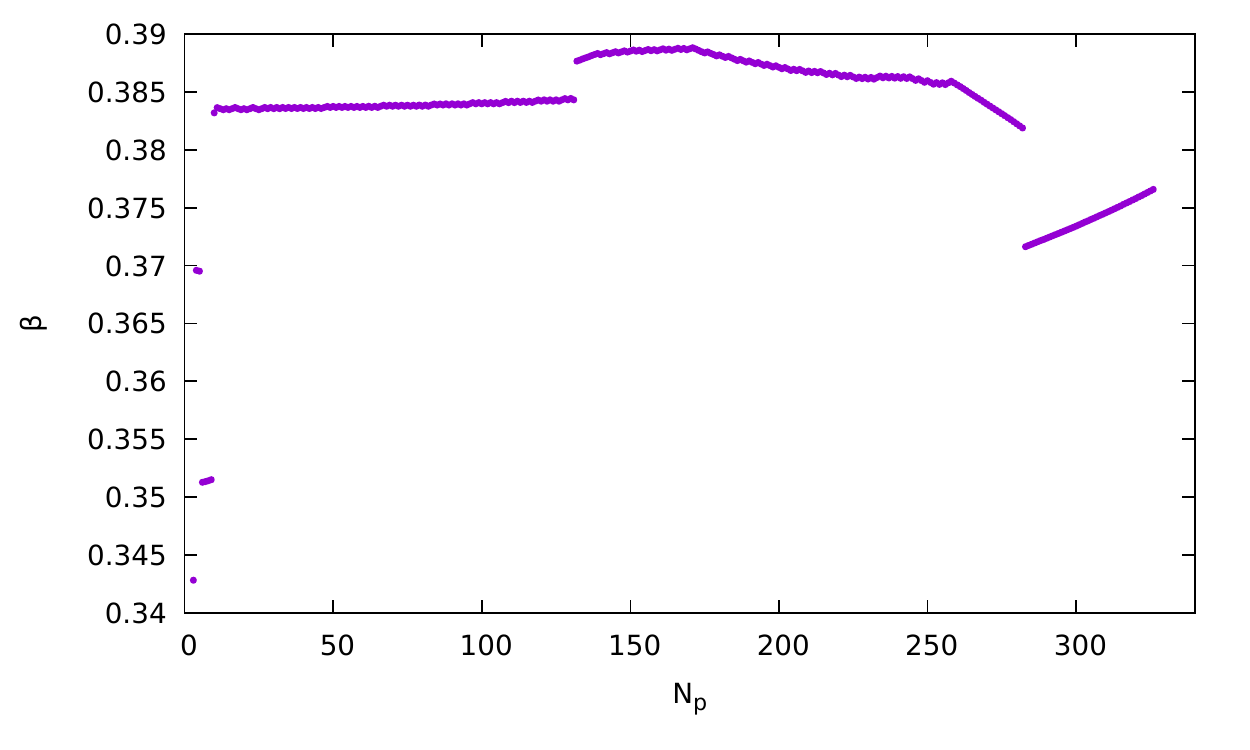}
        \end{minipage}%
        \begin{minipage}{0.5\textwidth}
            \centering
            \includegraphics[width=\linewidth]{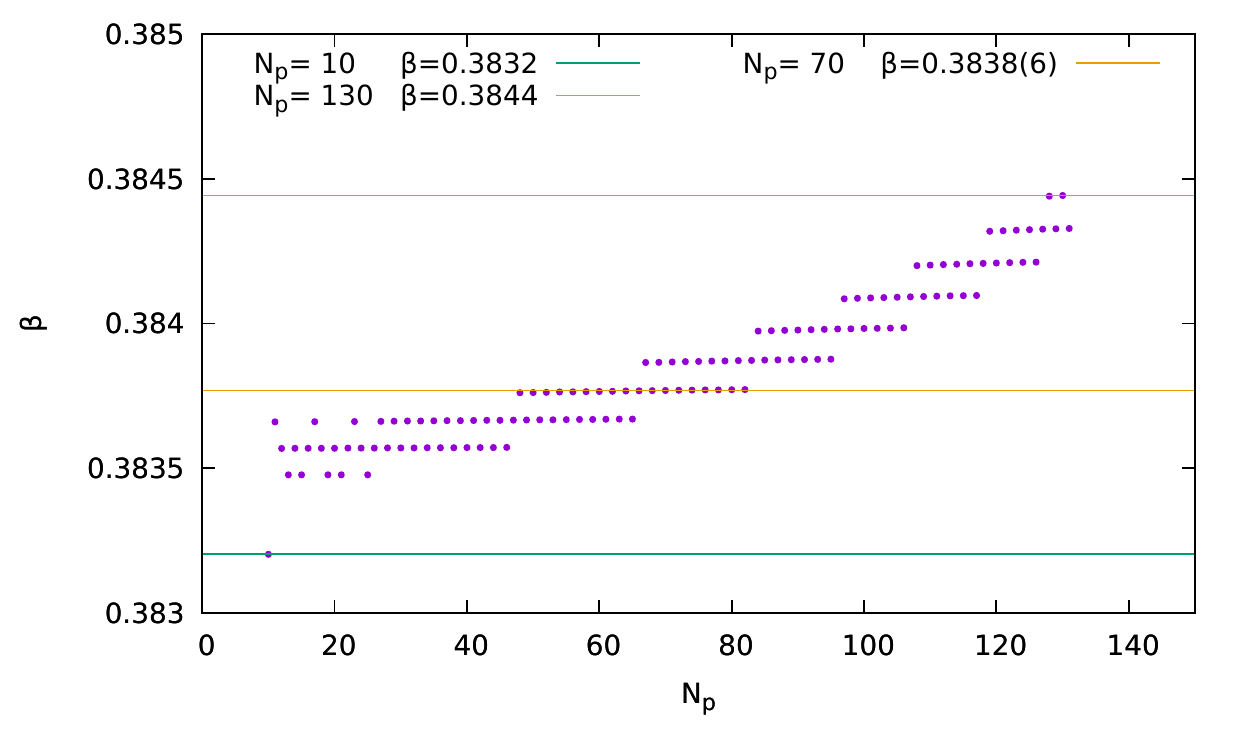}
        \end{minipage}
        \caption{Critical exponent $\beta$ as a function of the number of the most aligned points $N_p$ taken into consideration in the fit (left panel). The right panel is zooming in on the region where the value of the critical exponent is least dependent on $N_p$. In both plots $N = 3$ and $d = 3$. Other parameters of the calculation are the same as for \cref{scaling}.}
        \label{betap}
    \end{figure}
    As discussed in \cref{sec:IV}, a linear fit to the data is performed, the slope of which gives the critical exponents.
    This procedure inevitably leads to some error due to the fact that, as mentioned in \cref{sec:IV}, some points do not follow the critical scaling and thus deviate from the linear fit.
   Therefore, we need a criterion which enables us to select which points we should take into account for the fit.
    This is an important issue, since, as can be seen from \cref{nup,betap}, the value of a particular critical exponent and the relative fit error may change a lot if only a single data point is added to (or subtracted from) the fit.
    Since our goal is to extract the critical exponents from the scaling region, where we expect a linear behaviour of the observables (the logarithm of the \gls{ir} minimum and the logarithm of the curvature mass) as a function of $\ln \big| \varrho_0|_{t = 0} - \varrho^c_0 |_{t = 0} \big|$, it seems reasonable to use those consecutive points for the fit which exhibit the highest degree of collinearity. 
    As a criterion for collinearity, we choose the  Pearson correlation coefficient.
    However, the number of points which are taken into consideration still have a significant impact on the value of the critical exponents.
    This can be clearly seen from \cref{nup,betap}, where the critical exponents are shown as a function of the number of consecutive most aligned points.
    This is due to the fact that, as shown in \cref{al}, either the region where the consecutive most aligned points are contained moves while modifying the number of points, or it includes points in the fit which are increasingly further away from a straight fitting line.
    Thus, a criterion is still needed according to which one should select the number of consecutive collinear points.
    
    It is worth noting that the behavior seen in \cref{nup,betap} is not affected by varying the point density (number of  points per unit interval of the fitting region), indicating that the values of the critical exponents depend on the size of the fitting region rather than on the number of points contained in it.
    Ideally, if all points belong to the scaling region, all of them would exhibit perfect collinearity and thus the critical exponents would be completely independent of the fitting region.
    
    Our criterion is thus to identify the scaling region by searching for the range on the abscissa in \cref{nup,betap} where the critical exponent is least dependent on the number of consecutive collinear points and thus on the fitting region.
    Once this range is identified, we  assume that the actual value of the critical exponent lies somewhere between the highest and the lowest value of the critical exponent in that range.
    This gives a contribution to the total error on the critical exponent which is half the difference of the highest and the lowest value in that range.
    Finding such a range of consecutive collinear points is straightforward in the case of $\beta$:
    if the number of points in the range considered is too small, the range moves a lot along the abscissa when it is extended.
    Therefore, we discard the leftmost points in \cref{betap} (left panel).
    From a certain size onward the range does not move anymore, but merely grows within the critical region.
    However, if the range is too large, also points from outside the critical region are included, which is clearly seen by the jump in \cref{betap} (left panel) at $N_p \simeq 130$.
    Hence, this clearly restricts the size of the critical region.
    
    On the other hand, in the case of $\nu$, if the fitting region is too large, the exponent starts to change more rapidly, indicating that the corresponding fitting region includes points which do not follow a straight line in \cref{nu}. 
    Thus, we restricted the fit to a maximum value of $N_p \simeq 180$.
    However, from \cref{nup} it is clearly visible that it is much harder, if not almost impossible, to identify a critical region, because $\nu$ strongly depends on the fitting interval.
    
    We conclude this subsection by first noticing that the error originating from the choice of the fitting region is much larger than the error of the fit itself, i.e., the one extracted from the maximum deviation of the points in the fitting region from a straight line.
    Indeed, this error is at least one order of magnitude smaller than the  error from the choice of the fitting region.
    Second, as we discuss next, the error from choosing the correct fitting region is also much larger than the numerical error coming from the \gls{kt} scheme and the determination of the curvature mass, which is required for extracting $\nu$.
    For $\beta$, however, the discretization error which results in an uncertainty on the position of the minimum is again comparable to the error arising from the determination of the size of the critical region.

\subsection{Error on \texorpdfstring{$m^2$}{mass squared}}

    \begin{figure}
        \centering
        \begin{minipage}{0.5\textwidth}
            \centering
            \includegraphics[width=\linewidth]{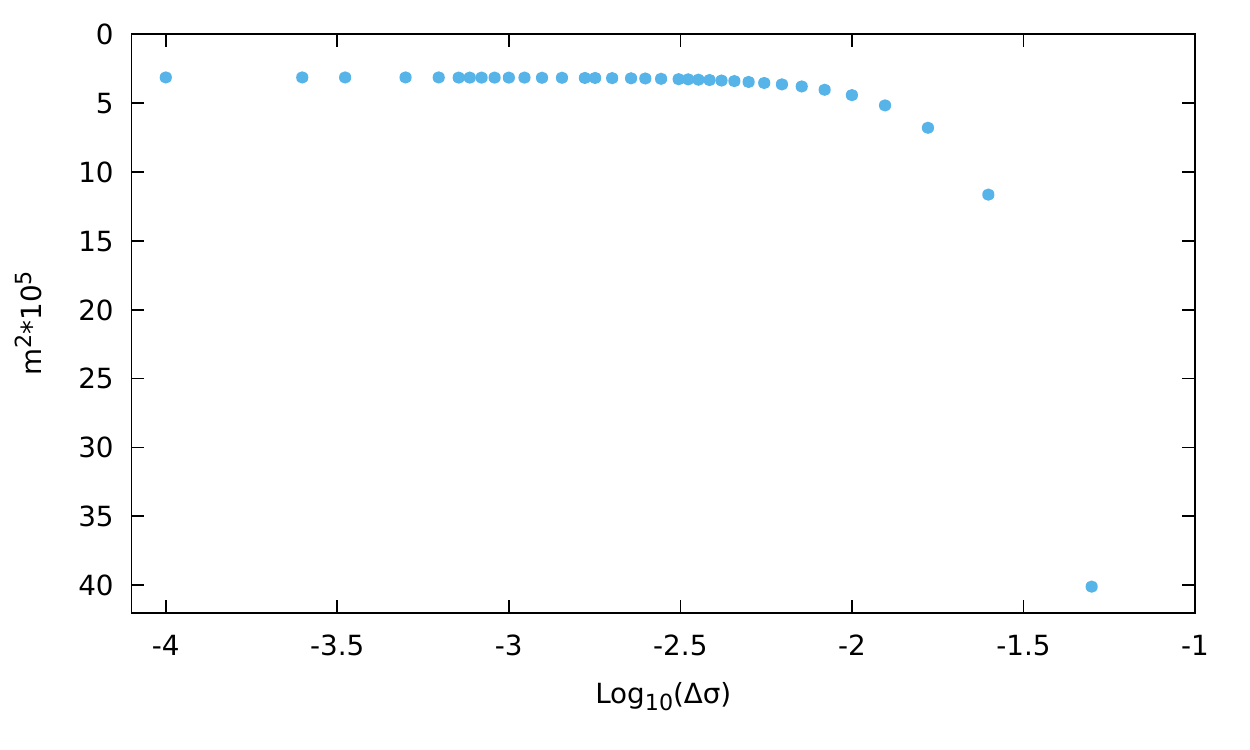}
        \end{minipage}%
        \begin{minipage}{0.5\textwidth}
            \centering
            \includegraphics[width=\linewidth]{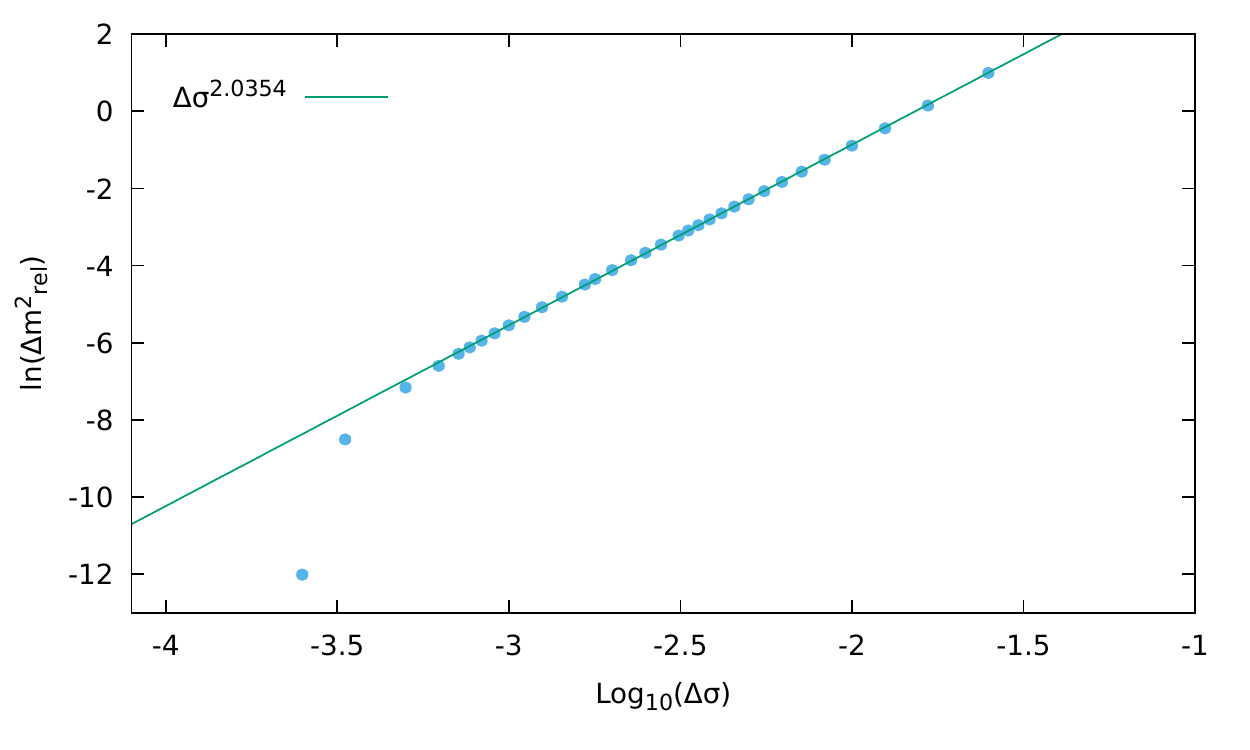}
        \end{minipage}
        \caption{Curvature mass squared in the \gls{ir}, $m^2$ (left panel), and relative deviation of $m^2$ from the value  calculated for the smallest cell size (right panel) as functions of the cell size $\Delta \sigma$.
        Here, $N=3$, $d=3$, $\Lambda = 1.0$, and $\sigma_{\text{max}} = 2.0$.}
        \label{m}
    \end{figure}

    In this subsection we analyze the impact of the discretization in field space, i.e., of the grid spacing $\Delta \sigma$, on the value of the renormalized mass $m^2$ and on the corresponding error.
    In \cref{m}, we show the curvature mass squared $m^2$ (left panel) and its relative deviation (right panel)
        \begin{align}
            \Delta m^2_{rel}(\Delta \sigma)=1 -\frac{m^2(\Delta \sigma)}{m^2(\Delta \sigma_{min})}
        \end{align}
    from the value $m^2 ( \Delta \sigma_{min} )$ calculated for the smallest cell size $\Delta \sigma_{min}$ (which in our case is $10^{-4}$) as a function of $\Delta \sigma$.
    Note that each point in these figures is the result of an \gls{frg} flow evolution, obtained for $\sigma_0 |_{t = 0} = 0.3$ in the symmetric phase ($\sigma_0^\ir = 0$) at final time $t_\ir = 25$ ($k_\ir \sim 10^{-11}$).
    The aim is to identify the value of $m^2$ from a plateau region, i.e., when its value does not change anymore upon changing the value of $\Delta \sigma$.
    From \cref{m} we observe that this happens for values of $\Delta \sigma \lesssim 10^{-2.5}$.
    Thus, in order to get a reasonable compromise between computational resources and needed precision, we use $\Delta \sigma = 0.0005$ (corresponding to the fourth point from the left in \cref{m} (left panel)).
    As one observes from the right panel of \cref{m}, this choice leads to a relative error in the determination of $m^2$ of the order of $\sim e^{-7} \simeq 10^{-3}$.
    
    As anticipated, the right panel provides an estimate of the order of magnitude of the relative errors, which is especially useful for those points which belong to the plateau in the left panel.
    The fact that these relative differences scale as $\Delta \sigma^\alpha$, with $\alpha>2$, is a direct consequence of the numerical scheme used, the \gls{kt} scheme, which has second-order precision in the spatial resolution.
    This means that it is possible to significantly reduce the error in the determination of $m^2$ by choosing a sufficiently small $\Delta \sigma$.
    We should also mention that there is no additional error from using a finite difference stencil in \cref{eq:28} to extract the curvature mass at $\sigma = 0$, for details see Refs.~\cite{Stoll:2021ori,Koenigstein:2021syz}.

    One final remark is related to the extrapolation of the value of $m^2$ in the \gls{ir}.
    As final value of $t$ in the \gls{ir} we have chosen $t = 25$, since at this point the \gls{frg} flow is effectively frozen in, i.e., no quantity changes anymore with \gls{rg} time.
    We convinced ourselves that this is true by computing $m^2$ for $t = 50$ and $t = 100$.
    The values of $m^2$ at these times were the same as at $t = 25$ up to machine precision.
    Thus, in the symmetric phase it is possible to reach arbitrarily small momentum scales to obtain \gls{ir} quantities.
    
\subsection{Error on \texorpdfstring{$\sigma_0^\ir$}{sigmazero in the IR}}

    Several aspects have to be taken into account regarding the determination of the value of $\sigma_0^\ir$.
    First of all, in the broken phase, $u ( t, \sigma )$ is negative for $\sigma < \sigma_0$ and approaches zero from below as $t \to \infty$.
    As a consequence, the denominator in \cref{eq:15} becomes very small when approaching the \gls{ir}, which leads to a stiff problem when solving the \gls{frg} flow equation numerically, i.e., the time step $\Delta t$ has to become increasingly smaller when $t \to \infty$ in order to avoid numerical instabilities.
    In practice, this prevents us to go arbitrarily far into the \gls{ir} within our setup.
    For recent advances on this issue, we refer to Ref.\ \cite{Ihssen:2023qaq}.
        \begin{figure}
            \begin{minipage}{0.5\textwidth}
                \centering
                \includegraphics[width=\linewidth]{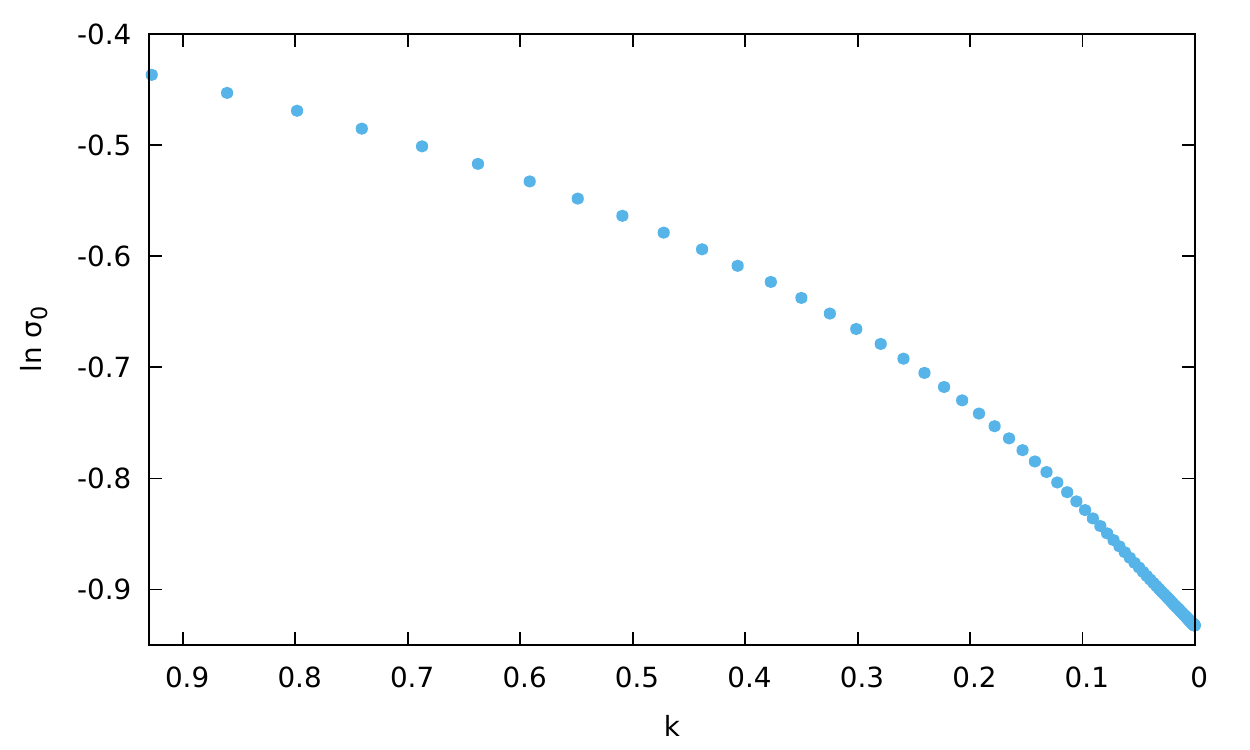}
            \end{minipage}%
            \begin{minipage}{0.5\textwidth}
                \centering
                \includegraphics[width=\linewidth]{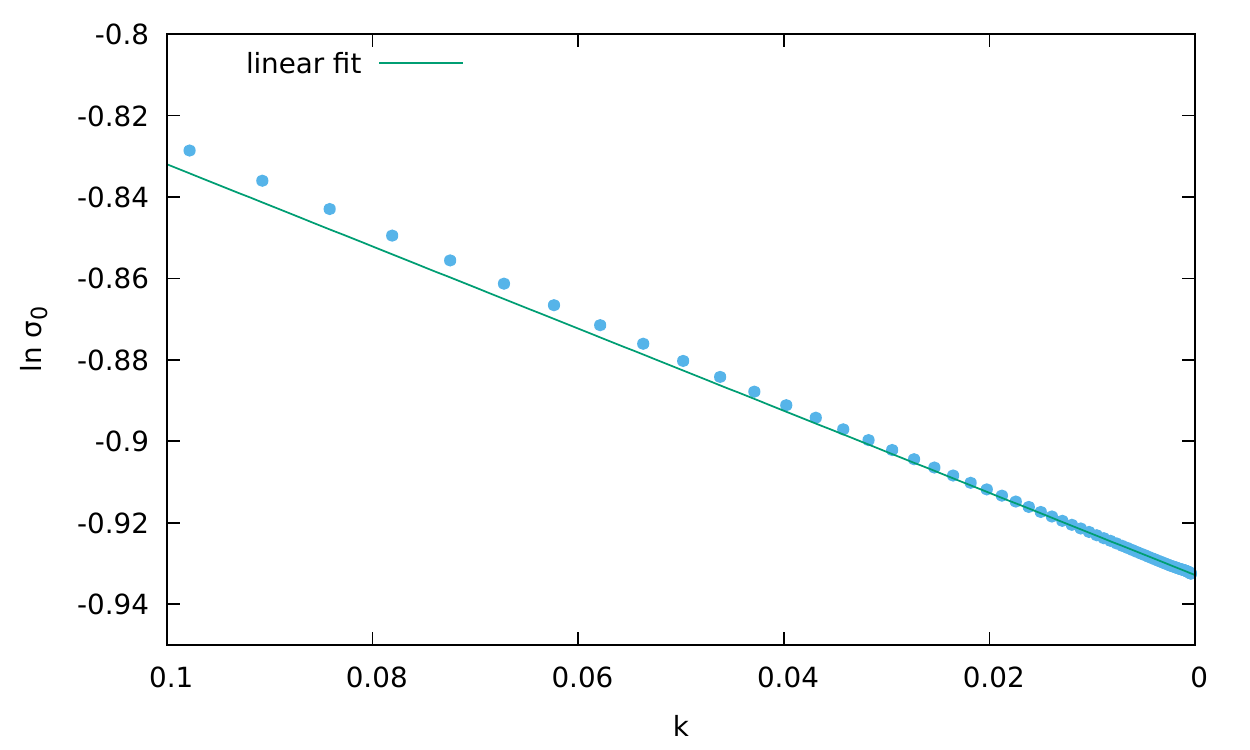}
            \end{minipage}
            \caption{Logarithm of the minimum $\sigma_0$ of the effective potential as a function of the momentum scale $k$, for $N = 3$ and $d=3$. 
            The right panel shows the range $0.1 \geq k \geq 0$ from the left panel in higher resolution, with the corresponding linear extrapolation to determine the minimum $\sigma_0^\ir$ at $k = 0$. Other parameters of the calculation are the same as for \cref{scaling}.}
            \label{fig:7}
        \end{figure}
    This problem is solved by determining the position of the minimum $\sigma_0^\ir$ via extrapolation.
    Here, we decided to perform an exponential extrapolation of $\sigma_0$ as a function of $k$ via a linear extrapolation of the values of $\ln \sigma_0$
        \begin{align}
        	&  \ln \sigma_0 = a \, k + b \, ,  &&  \Longleftrightarrow &&  \sigma_0 = e^b \, e^{a k} \, ,
        \end{align}
    as can be seen from \cref{fig:7}.
        \begin{figure}
            \begin{minipage}{0.5\textwidth}
                \centering
                \includegraphics[width=\linewidth]{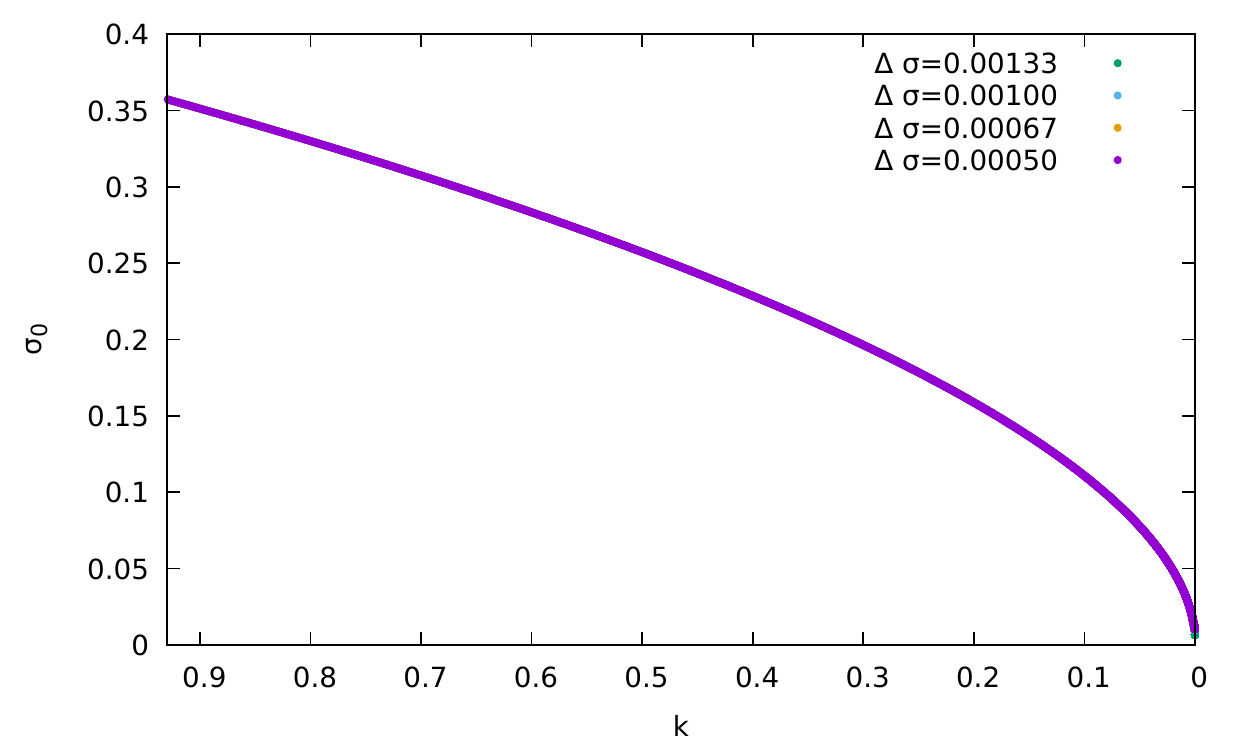}
            \end{minipage}%
            \begin{minipage}{0.5\textwidth}
                \centering
                \includegraphics[width=\linewidth]{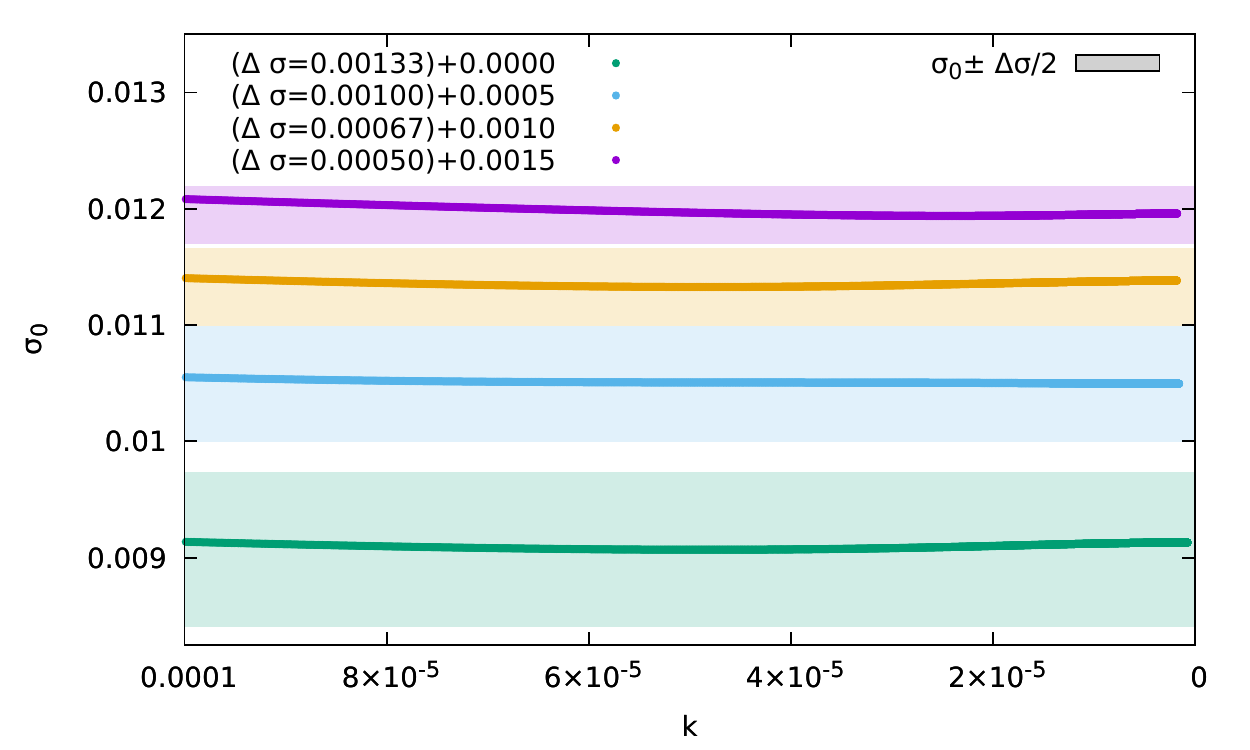}
            \end{minipage}
            \caption{Position of the minimum of the effective potential $\sigma_0$ as a function of the momentum scale $k$ for different values of the cell size $\Delta \sigma$, for $N = 3$ and $d=3$.
            The right panel shows the range $0.0001 \geq k \geq 0$ from the left panel in higher resolution, where the fluctuations in the value of $\sigma_0$ can be seen.
            Each band around a line has a width $\Delta \sigma$.
            In order to avoid overlapping bands, we have performed a global shift of $\sigma_0$ for the cases $\Delta \sigma = 0.001$, $0.00067$, and $0.0005$ by $0.0005$, $0.001$, and $0.0015$, respectively.
            Other parameters of the calculation are the same as for \cref{scaling}.}
            \label{fig:8}
        \end{figure}
    This functional form of the extrapolation is motivated by the behavior of $\sigma_0$ in the large-$N$ limit, where such an exponential scaling is exact, as can be shown by solving the \gls{frg} flow equation analytically via the method of characteristics (see Refs.\ \cite{Grossi:2019urj,Thomas:99}).
    This extrapolation is the reason for the small oscillations of $\ln \sigma_0^\ir$ observed in \cref{beta}.
    
    Even if the extrapolation procedure were to give the correct value, the uncertainty on $\sigma_0^\ir$ is still limited from below by the grid spacing $\Delta \sigma$.
    Assuming that the correct value lies within a cell, one can still determine whether $\sigma_0$ lies to the left or the right of the cell center, such that the uncertainty is $\Delta \sigma_0^\ir \lesssim \Delta \sigma/2$.
    
    Indeed, it can be seen from \cref{fig:8} that, starting with the same fixed $\sigma_0^{\uv}$, the position of the minimum is independent of the cell size $\Delta \sigma$ for larger momentum scales and becomes dependent on $\Delta \sigma$ as $k$ decreases (the shifts of the curves in the right panel, which are performed to increase the visibility of the $\Delta \sigma$ bands, are smaller than the actual differences between the curves).
    However, we observe that, for any given $\Delta \sigma$, for small $k$ the fluctuations of $\sigma_0$ as a function of $k$ stay within a band given by $\pm \Delta \sigma/2$, confirming the aforementioned assumption for the error on $\sigma_0^\ir$.
    In addition, these fluctuations seem to decrease as $\Delta \sigma$ is reduced.
    
    Thus, assuming $\Delta \sigma/2$ as the error on every value of  $\sigma_0^\ir$,  each point in \cref{beta} has an error given by
        \begin{equation}\label{eq:31}
            \delta \ln \sigma_0^\ir = \frac{\delta \sigma_0^\ir}{\sigma_0^\ir} = \frac{\Delta \sigma}{2\sigma_0^\ir} \, .
        \end{equation}
    Note that this error depends on $\sigma_0^\ir$ itself. 
    We then consider the value of $\beta$ extracted from the line which passes through the points $\Big((\varrho_0|_{t = 0} - \varrho^c_0 |_{t = 0})^{left}, \ln \sigma_0^{\ir, left}- \delta\,\ln \sigma_0^{\ir, left}\Big)$ and $\Big((\varrho_0|_{t = 0} - \varrho^c_0 |_{t = 0})^{right}, \ln \sigma_0^{\ir, right}+\delta\,\ln \sigma_0^{\ir, right}\Big)$, where $\, \delta  \ln \sigma_0^{\ir, left} \mbox{ and }\, \delta  \ln \sigma_0^{\ir, right}$ are the errors of the leftmost and rightmost points of the selected fit interval.
    The difference of this value of $\beta$ and the one extracted from the slope of the straight line in \cref{beta} gives our estimate for $\delta \beta$, which then reads
        \begin{equation}\label{eq:32}
            \delta \beta = \frac{\, \delta  \ln \sigma_0^{\ir, left}+\, \delta  \ln \sigma_0^{\ir, right}}{\Delta \varrho_{fit}}  \; ,
        \end{equation}
         where $\Delta \varrho_{fit}$ is the range in $\ln (\varrho_0|_{t = 0} - \varrho^c_0 |_{t = 0})$ where the linear fit is performed.
        
    Using typical values encountered during the calculations, i.e., $\Delta \varrho_{fit}\simeq 3.5$, $\Delta \sigma=0.0005$  one finds
        \begin{align}
           \delta \beta \simeq 0.006 \, .
        \end{align}
    We will assume this as error on the critical exponent $\beta$.
    It is worth noticing that this contribution is significantly  larger than the one coming from the fit error, i.e., the one extracted from the maximum deviation of the points from the straight-line fit, which is usually of order
        \begin{align}
           \delta \beta_{fit} \sim 0.0002 \, .
        \end{align}
        If one wants to have an upper and lower bound for the error, one can use the minimum and the maximum value for $\sigma_0^\ir$ in the fit region in \cref{eq:32}.
   One gets
             \begin{align}
              \delta \beta_{>} = 0.024 \qquad \delta \beta_{<} = 0.007 \, .
            \end{align}     
             Thus, our estimate of the error seems a good compromise between the highest and the lowest possible errors.

\bibliography{Critical_exponents_ON_FRG}

\end{document}